\documentclass[traditabstract]{aa}

\usepackage{amsmath,amssymb}
\usepackage{graphicx,epsfig,subfigure}
\usepackage{xspace}
\usepackage{hyperref}
\usepackage{aas_macros}
\usepackage{color}
\usepackage{mathrsfs}
\usepackage{txfonts}
\usepackage{natbib}
\bibpunct{(}{)}{;}{a}{}{,}
\def\etal{{\it et al.,}\xspace}

\def\ie{{\it i.e.}\xspace}
\def\eg{{\it e.g.}\xspace}

\def\wrt{with respect to\xspace}

\def\to                 {\ensuremath{\rightarrow}\xspace}
\newcommand{\pl}{\textsf{PLANCK}\xspace}
\newcommand{\wmap}{\textsf{WMAP}\xspace}
\newcommand{\hp}{\texttt{HealPix}\xspace}
\newcommand{\ns}{\ensuremath{n_\text{side}}\xspace}
\newcommand{\lmax}{\ensuremath{\ell_\text{max}}\xspace}
\newcommand{\cl}{\ensuremath{C_\ell}\xspace}
\providecommand{\E}[1]{\langle#1\rangle}

\providecommand{\norm}[1]{\lvert#1\rvert}

\newcommand{\intphi}{\ensuremath{\int \frac{d\phi}{2\pi}}\xspace}

\def\deg{\degr}
\newcommand{\mpa}{\texttt{multi-patch}\xspace}
\newcommand{\inp}{\texttt{FS-inpainting}\xspace}
\newcommand{\ft}{\texttt{FourierToeplitz}\xspace}
\newcommand{\ghz}{GHz\xspace}
\newcommand{\kmod}{\ensuremath{\lvert \vec k \rvert}\xspace}
\newcommand{\refeq}[1]{Eq. (\ref{eq:#1})\xspace}
\newcommand{\fsky}{\ensuremath{f_\text{sky}}}
\providecommand{\norm}[1]{\lVert#1\rVert}
\providecommand{\E}[1]{\langle#1\rangle}

\newcommand{\Cld}{\ensuremath{C_\ell^{\text{d}}}\xspace}
\newcommand{\Nz}{\ensuremath{N^{(0)}}\xspace}
\newcommand{\Nlz}{\ensuremath{N_\ell^{(0)}}\xspace}
\newcommand{\Nkz}{\ensuremath{N_K^{(0)}}\xspace}
\newcommand{\None}{\ensuremath{N^{(1)}}\xspace}
\newcommand{\Nlone}{\ensuremath{N_\ell^{(1)}}\xspace}
\newcommand{\Nkone}{\ensuremath{N_K^{(1)}}\xspace}
\newcommand{\Ntwo}{\ensuremath{N^{(2)}}\xspace}
\newcommand{\Nltwo}{\ensuremath{N_\ell^{(2)}}\xspace}
\newcommand{\Nktwo}{\ensuremath{N_K^{(2)}}\xspace}
\newcommand{\Nmc}{\ensuremath{N^\text{MC}}\xspace}
\newcommand{\Nlmc}{\ensuremath{N_\ell^\text{MC}}\xspace}
\newcommand{\Nkmc}{\ensuremath{N_K^\text{MC}}\xspace}

\newcommand{\sect}{Sect.\xspace}

\newcommand{\wwwhealpix}{\url{http://healpix.jpl.nasa.gov}}
\newcommand{\wwwlenspix}{\url{http://cosmologist.info/lenspix}}
\newcommand{\wwwcamb}{\url{http://camb.info}}
\newcommand{\wwwfastlens}{\url{http://irfu.cea.fr/Ast/fastlens.software.php}}
\newcommand{\wwwmrs}{\url{http://jstarck.free.fr/mrs.html}}

\definecolor{gris}{gray}{0.50}

\begin{document}

\title{A hybrid approach to cosmic microwave background lensing
  reconstruction from all-sky intensity maps}
\author{ S. Plaszczynski\inst{1}
        \and
        A. Lavabre\inst{1}
        \and
        L. Perotto\inst{2}
        \and
        J.-L. Starck\inst{3}
        }

\institute{
           Laboratoire de l'Acc\'el\'erateur Lin\'eaire (LAL), 
Univ Paris-Sud, CNRS/IN2P3, Orsay, France.\\
            \email{plaszczy@lal.in2p3.fr,lavabre@lal.in2p3.fr}
        \and
Laboratoire de Physique Corpusculaire et de Cosmologie (LPSC), Universit\'e Joseph Fourier Grenoble 1, CNRS/IN2P3, Institut Polytechnique de Grenoble, 53, rue des Martyrs, 38026 Grenoble Cedex, France\\
           \email{perotto@lpsc.in2p3.fr}
        \and
Laboratoire AIM (UMR 7158), CEA/DSM-CNRS-Universit\'e Paris Diderot,   
IRFU, SEDI-SAP, Service d'Astrophysique,  Centre de Saclay, 
F-91191 Gif-Sur-Yvette cedex, France \\
            \email{jstarck@cea.fr}
} 
\date{Received \today}

\abstract{On the basis of realistic simulations, we propose a hybrid method to reconstruct the
  lensing potential power spectrum, directly on \pl-like cosmic microwave background frequency maps. 
  This involves the use of a large Galactic mask and the treatment of strong
  inhomogeneous noise. For $\ell \lesssim 100$, we
  show that a full-sky inpainting method, which was previously described,
  still allows a minimal variance reconstruction, with a bias that
  must be accounted for by a Monte Carlo method 
  but that does not couple to the deflection field. For $\ell \gtrsim 100$, we develop a method based on tiling
  the cut-sky with local $10\deg \times 10\deg$ overlapping tangent
  planes (referred to in the following as {\it patches}). We tackle
  various issues related to their size/position, non-periodic boundaries, and
  irregularly sampled data of the planes after the sphere-to-plane projection. We
  show that the predominant noise term of the quadratic lensing estimator 
  determined from an apodized patch can still be recovered 
  directly from the data. To prevent any loss of spatial accuracy, we
  developed a tool that allows the efficient determination of the complex Fourier
  series coefficients from a bi-dimensional irregularly sampled
  dataset, without performing any interpolation. We show that
  our multi-patch approach enables the lensing power spectrum to be reconstructed
  with a very small bias, thanks to the omission of a Galactic 
mask and smaller noise inhomogeneities, as well as an almost
minimal variance. At each stage the data quality can be
assessed and simple bi-dimensional spectra compiled, which
allows the control of local systematic errors.
}


\authorrunning{S. Plaszczynski \etal}

\titlerunning{A hybrid approach to CMB lensing reconstruction}

\maketitle  

\section*{Introduction}

Experiments have now reached the sensitivity in terms of both resolution and
noise, to detect the tiny deflection of the cosmic microwave
background (CMB) photons ($\sigma_d \simeq 2.7\arcmin$) 
by the irregular distribution of matter, in their journey from the last scattering
surface to Earth. First results on the power spectrum of this deflection field
have been reported by the \texttt{ACT} \citep{act} and \texttt{SPT} \citep{sptlens}
collaborations. The \pl spatial mission should soon provide firm measurements.
This information provides access to a new cosmological observable that
is sensitive to an epoch ($1 \lesssim z \lesssim 3$) much more recent than the CMB
decoupling one ($z \simeq 1100$), giving us a lever-arm to lift the
so-called \emph{geometrical degeneracy} \citep{radek}, but using one single consistent
data-set. In particular, it probes the  matter density fluctuations,
on scales where the free-streaming of massive neutrinos significantly erases the
power spectrum of these fluctuations \citep{pastor}, and is expected
to help us to determine their total mass by means of global cosmological fits.

On statistical grounds, the properties of the (nearly) optimal quadratic
estimator for lensing power reconstruction are now well-understood, both in
the (infinite) flat sky limit and across the complete sphere \citep{hu2,hu3}.

However, real data are affected by contaminants, mostly
Galactic dust and point sources in the case of CMB frequency maps,
requiring a revised means of lensing reconstruction on a
cut-sky, which is a non-trivial task. In addition, the scanning strategy of the
specific instrument, particularly in the case of \pl, induces some strong spatial-noise
inhomogeneities that are not taken into account in the classical
estimator of lensing,  and must be corrected for by Monte Carlo simulations.  
In the general case, both effects cannot be distinguished during the reconstruction process.

In a previous study \citep{papier1}, we optimized a sparse inpainting
procedure to restore the missing data inside the mask, without
significantly biasing the lensing results.
We however neglected the noise inhomogeneity. Furthermore, the
work was oriented towards component-separated maps, so that the mask to be filled was rather small (about 10\% of the sky).

However before having to adopt a component separation method that mixes different
maps, we wish to investigate in this paper whether the lensing
potential can be
reconstructed more directly in individual intensity CMB maps, which is
indeed a necessary
step in assessing the possible systematic errors. 
For \pl, the channels under consideration correspond to 100, 143, and 217\ghz \citep{bluebook}.
This requires the treatment of much larger masks. We will also
consider the strong spatial-noise inhomogeneities induced by the
scanning strategy.

We revisit the sparse inpainting method in this new configuration (a
$30\%$ mask + inhomogeneous noise) and show that  i) the
estimator under these conditions is strongly biased and ii) a Monte Carlo approach 
can be used to correct for this bias.
We also propose an alternative method (\mpa)
that allows us to completely avoid the Galactic region, during the
development of which we solved a number of issues related to the pixelized-sphere to plane projection.

After rapidly reviewing the various noise contributions to the
quadratic estimator (QE) in \sect \ref{sec:ideal}, and the common
simulations used in \sect \ref{sec:sim}, we update our full-sky
inpainting analysis (hereafter denoted \inp) in \sect \ref{sec:inp}. 
Most of the paper in \sect \ref{sec:mpa}, then deals with
resolving issues related to the projection of a non-periodic signal
from a pixelized sky onto a local patch. 
In particular, we present a new algorithm (detailed in the
Appendix) that allows a fast reconstruction of band-limited Fourier series
coefficients from irregularly sampled data, without performing any interpolation. 
We then compare both methods, optimize the results in \sect
\ref{sec:results}, and argue that a hybrid reconstruction is the most
appropriate. In this hybrid approach, the full sky lensing reconstruction presented in  \citet{papier1} is used at low $\ell$, 
with an additive Monte Carlo bias correction, while at high $\ell$, the new \mpa method is advocated.

This method allows the direct reconstruction of lensing signal from \pl-like CMB
frequency maps (namely those at the 100, 143, and 217 \ghz).
While it would be premature to decide today
whether performing a multi-map component separation
provides a more accurate recovery of the lensing signal, we give some elements of the discussion in the conclusion.

\section{A brief review of the quadratic estimator}
\label{sec:ideal}

The gravitational lensing potential $\phi$ is a scalar isotropic field
\citep[for a review, see \eg][]{challinor}
that spatially \textit{remaps} the CMB photons according to 

\begin{equation}
  \label{eq:remap}
  T(\vec n)=T_\text{CMB}(\vec n +\vec d(\vec n)),
\end{equation}
where $\vec d=\nabla \phi$ is the deflection field, which has a power
spectrum on the sky $C_\ell^d=\ell(\ell+1)C_\ell^\phi$, or
$C_K^d=k^2 C_K^\phi$ in the flat sky limit.
This process slightly breaks the Gaussianity of the CMB field, and estimators have been searched for
in order to extract the lensing information using its very local properties

The quadratic estimator was proposed by \citet{hu2}. For CMB temperature
anisotropies, it consists in taking the
(weighted) convolution of the observed Fourier modes
according to 
\begin{equation}
\label{eq:esti}
  \hat \phi(\vec K)= A_K\int \frac{d^2 k_1}{(2\pi)^2} T(\vec k_1)T(\vec K-\vec k_1)F(\vec k_1,\vec K-\vec k_1) ,
\end{equation}
where the normalization $A_K$ and filter $F$ are determined
by requiring the estimator to be un-biased and have a minimum variance at the leading noise order  (so-called \Nz). For an idealized experiment, one gets
\begin{eqnarray}
\label{eq:esti2}
A_K&=& \left(\int \frac{d^2 k_1}{(2\pi)^2} f(\vec k_1,\vec K-\vec k_1) F(\vec k_1,\vec K-\vec k_1) \right)^{-1}  \nonumber \\
F(\vec k_1,\vec k_2)&=&\dfrac{f(\vec k_1,\vec k_2)}{2{C_{k_1}}^{\text{tot}}{C_{k_2}}^{\text{tot}}} \\
\text{where}~f(\vec k_1,\vec k_2)&=&(\vec k_1+\vec k_2)\cdot \vec k_1 {\tilde C_{k_1}}+(\vec k_1+\vec k_2)\cdot \vec k_2 { \tilde C_{k_2}}\nonumber 
\end{eqnarray}
The filter $F$ involves on the numerator the CMB "true" unlensed
power-spectrum ($\tilde C_k$) , and on the denominator the "observed"
one $C_k^{\text{tot}}$, which is assumed to be a pure beam-deconvolved
Gaussian signal with un-coupled homogeneous noise.

Since this estimator involves only simple operations, it is computable in a
few minutes on any standard computer.
Its generalization to spherical harmonics across the full-sky was
performed in \cite{hu3}.

The full likelihood estimator was developed by \citet{hirata1},
who demonstrated that it gives results very close to the quadratic
one, given the current noise level, but involves much heavier
computations.

The covariance of the $\hat \phi$ estimator is  
related to the true lensing potential spectrum $C_k^\phi$ through
\begin{equation}
  K^2 {\E{{\hat \phi(\vec K)}^*{\hat \phi(\vec K')}}}= (2\pi)^2
  \delta(\vec K - \vec K') [ K^2 C_K^\phi +{\Nkz + \mathscr{O}(C_K^\phi)}]
\end{equation}
and remarkably, the
noise term is directly related to the estimator normalization \refeq{esti2}
\begin{equation}
  \label{eq:n0}
  \Nkz=K^2 A_K .
\end{equation}
This corresponds to the \emph{Gaussian} term, in the sense that it
comes from the standard \textit{disconnected} part of the four-point
correlator that appears when computing the noise and is therefore decoupled from the $\phi$
field. Equivalently, it represents the
power of the QE when running it on unlensed maps.
 
A first-order power-spectrum correction term was soon afterwards discovered by \cite{kesden}. 
This comes from the \textit{connected} part of the correlator, and
hence depends on the $\phi$ field itself $\None(\phi)$.

When actually coding the estimator for the \pl experiment, we still
noted a poorly understood bias at low $\ell's$ which was finally
identified by \citet{duncan} as a non-negligible second-order
term that can be estimated analytically and was called $\Ntwo(\phi)$. 
Another way of taking this noise into account is
to use a simple "trick" proposed by P. Bielewicz \citep{duncan}, which consists in inserting
the \textit{lensed} spectrum 
into the numerator of \refeq{esti2}. This approach was shown to capture even more precisely 
the second-order contribution than the $\Ntwo$ but slightly increases the error in the reconstructed signal.

This is not however the end of the story. Our simulations did
not initially incorporate the spatial inhomogeneity of the
noise, owing to the \pl scanning strategy.
This strategy induces correlations between the different Fourier modes
leading to spurious signal reconstruction in the QE. It was shown in
\citet{inhom} that the noise inhomogeneity also affects the QE expectation value 
resulting in a low-$\ell$ bias in the power spectrum that can be
analytically estimated under the white noise hypothesis. 
However, this \emph{mean field} approach still misses another
$\Nkone$-like term, which is non-computable analytically but affects the
whole lensing spectrum.

Finally, owing to the foreground signal, one can never experimentally
make use of the signal across the full sphere.
In  this case, the spherical harmonics no longer form a "natural" basis and the issue of building a
good estimator for lensing is non-trivial. 
Even the inverse variance weighting of the map (\eg \citep{dore}),
which is a computationally
very challenging task and sometimes referred to as being "optimal",
does not provide an unbiased estimate of the lensing spectrum, because
of the large mode coupling introduced by the mask and the inhomogeneous noise.

To take into account these last two effects, namely the treatment of the
masked region and inhomogeneous noise, we add to the estimator
covariance a $\Nkmc(\phi)$ term that can in general depend on
the lensing field.

In summary, the deflection estimate variance from applying the QE to 
data using a given method, includes the following terms:
\begin{equation}
  \label{eq:varphi}
  \begin{split}
  K^2 {\E{{\hat \phi(\vec K)}^*{\hat \phi(\vec K')}}} &= (2\pi)^2
  \delta(\vec K - \vec K') \\
  & \cdot [C_K^d +\Nkz+\Nkone+\Nktwo + \Nkmc],
  \end{split}
\end{equation}
where 
\begin{itemize}
\item $C_K^d$ is the sought deflection spectrum;
\item \Nkz is determined on the data given the knowledge of the;
  underlying true power spectrum
\item \Nkone, and \Nktwo can be computed analytically. Since they depend
  on the searched field, one may need to setup an iterative
  determination. In our simulation, we simply use the
  true deflection spectrum from our test
    cosmology (\sect \ref{sec:sim}) to compute them.
\item \Nkmc is the bias of the power spectrum estimator, which depends on the
  inhomogeneous properties of the noise and the way in which we deal with the
  Galactic contamination (and the coupling of both).
\end{itemize}

The desired properties of \Nkmc are to have small value,
while still keeping the optimal variance for the estimator, and decoupled from the
lensing field. In the following, we study this term in two methods using a set
of simulations with inhomogeneous noise and a large mask.

We somewhat loosely switch to the multipole notation ($\ell$) in the
full-sky case, the formal connection being performed in
\eg \citet{hu1}. We recall that on a square patch of size $L\times L$, the discrete Fourier
modes are located on a grid 
\begin{equation}
  \label{eq:grid}
 \vec{k}=k_{i,j}= \Delta k \left( \begin{array}{c}
i\\
j\\\end{array} \right)  
\end{equation}
with $\Delta k=\tfrac{2\pi}{L} ~(\simeq 35$ for $L=10\deg$), and $(i,j)$
being integers. In these units, the power spectrum $C_{\kmod}$ is
equivalent to \cl in the flat sky limit \citep{white}.

\section{Simulations}
\label{sec:sim}

To evaluate the performance of our algorithms, we produced a set of
realistic \pl-like frequency maps, \ie a combination of all channels of a
given frequency. The experimental characteristics are the ones
published in \citet{hfi} corresponding to almost ten months of data-taking.

The most interesting channels for CMB analyses using \pl data, are the 100,
143 and 217\ghz ones, where the Galactic dust contamination increases with
frequency but is still sub-dominant and other Galactic foreground
types of emission (such as synchrotron or free-free) which decrease with the
frequency band, are weaker than the CMB \citep{bluebook}. The resolution of the instrument,
which is crucial to the lensing reconstruction, goes however in the other
direction with an average full width at half maximum of the scanning beams of about
$9.5\arcmin, 7.1\arcmin,$ and 4.7 $\arcmin$,  respectively \citep{hfi}.
We chose to focus on the 217 \ghz channel, because it is the
most challenging for lensing, requiring the largest Galactic mask.

We thus developed the following pipeline for our simulations.

We start with a $\Lambda$CDM cosmology \{$\text{H}_0=72$, $\Omega_\text{b}h^2=0.023$, $\Omega_{\text{CDM}}h^2=0.11$, $Y_{\text{He}}=0.24$, $N_{\text{eff}}=3.04$, $\tau = 0.09$, $n_\text{s}=0.96$, $A_\text{s}=2.4 \times 10^{-9}$\}, which is consistent with the \wmap seven-year best-fit model \citep{wmap7},
and run the Boltzmann code \textsf{CAMB} \footnote{\wwwcamb} to produce
the corresponding spectra of CMB intensity/polarization anisotropies and lensing
potential, using \textsf{Halofit} \citep{halofit} for non-linear corrections. Both lensed/unlensed spectra
are computed with the code. In the following, we focus on
temperature maps since this is the best-suited observable for
reconstructing lensing in a \pl-like case. In the
following we denote as "fiducial" this true deflection spectrum.

These spectra then feed the \textsf{LensPix}\footnote{\wwwlenspix} code, which provides a
full-sky high resolution map in the \hp\footnote{\wwwhealpix} scheme
(\ns=2048). We use a cutoff
\lmax=3000. We verified that the resultant lensed spectrum is in excellent
agreement with the theoretical ones, up to
$\ell \lesssim 2750$, which is largely sufficient for our analysis.
One hundred realizations of these maps were produced.
We refer to these maps, which are assumed to represent the data, as the
$H_1$ set (\ie lensed).

Starting from the \textsf{CAMB} lensed power-spectra, we also produced
one hundred Gaussian realizations using the standard \hp tools (namely
\verb-syn_alm_cxx/alm2map_cxx-) which help us in de-biasing the
lensing power spectrum estimate. In the following we will refer to these maps as the $H_0$ set (\ie unlensed).

Maps were then all smoothed by a $4.7\arcmin$ Gaussian beam
using \hp standard tools.

The Gaussian correlated noise in the maps is generated according to its 
spectrum measured in \citet{hfi}. More precisely, the real and
imaginary parts of the spherical harmonics coefficients $a_{l,m}$ are randomly
drawn from an independent Gaussian distribution with zero mean and a
variance given by the measured spectrum $N_l$  using the standard \hp
\texttt{syn\_alm\_cxx} procedure.
We then transform the coefficients into real
space, using the \texttt{alm2map\_cxx} procedure. Each pixel value in the map
is then weighted according to the square-root of the number of hits
in that pixel, preserving the total variance. One hundred of these maps, each with a different seed for the phases, 
were produced and added to the signal maps. 

We then apply a $30\%$ mask obtained by smoothing a higher
  resolution (857\ghz) map to avoid the leading dust contamination in the Galactic plane.

Since it is not the scope of this study to investigate the systematics
errors caused by point-source residuals, we assume that a point-source mask
with perfect completeness is available. We produce this by
combining the \pl Early Release Compact Source Catalog of point
sources detected in the $143$, $217$, and $353~\ghz$ channels and including Sunyaev-Zel'dovich clusters (ESZ) and dust cold cores (CC) 
\citep{ERSC,ESZ,ECC}, the \wmap seven-year catalog of point sources with a positive flux in the W band \citep{wmapps}, and a catalog of IRAS/2MASS IR sources whose flux at $100~\mu m$ is above $2~\text{Jy}$ \citep{iras,2mass}. Each catalog entry is masked by a $5 \sigma \simeq 10\arcmin$ radius disk.
This mask covers $\simeq 1.7\%$ of the sky out of the Galactic plane. 
When combining it with the Galactic one, we are left with a fraction $\fsky=0.69$ of the sky. 

\begin{figure}[htbp]
  \centering
  \includegraphics[width=.3\textwidth,angle=90]{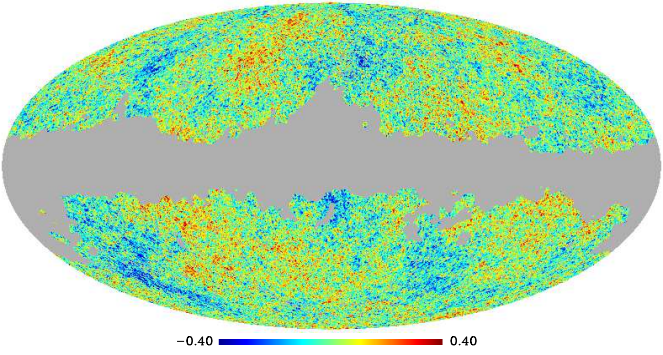}
  \caption{\label{fig:ex_simu} Example of one of our simulated lensed
    temperature map,
    using the procedure described in the text. Units are
    $mK_\text{CMB}$. The gray region corresponds to the Galactic mask
    we propose to use. A point-source mask is also included, but barely
  visible, being more clearly seen on Fig \ref{fig:patch_inp}.}
\end{figure}

Figure \ref{fig:ex_simu} shows one of these simulated maps.

\section{Update on the \inp method}
\label{sec:inp}

In \citet{papier1}, we studied the impact of an inpainting method
to fill in, with an appropriate statistical mixture, a rather small mask
(cutting a $\simeq 10\%$ fraction of the sky). It was oriented toward component-separated maps,
where such a level of final masking is to be expected.
Here, we push the algorithm further to its limits by studying the  
filling of the large mask defined in \sect \ref{sec:sim}
($\simeq 30\%$ of the sky). Furthermore, we add spatially inhomogeneous
noise, which most certainly affects the results of the algorithm. 

Among the inpainting algorithm implemented within the Multi-Resolution 
on the Sphere (MRS) package\footnote{\wwwmrs}, we found that the most
robust results are obtained with the spherical harmonics $L_1$ norm
minimization using wavelet packet variance regularization \citep{inp1,inp2}.

\begin{figure}[htbp]
  \centering
  \includegraphics[width=.3\textwidth,angle=90]{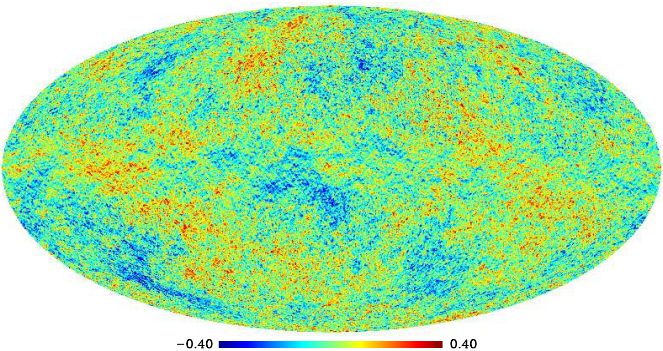}
  \caption{\label{fig:ex_simu_inp} Inpainted map corresponding
    to filling the Galactic+point-source mask of Fig \ref{fig:ex_simu}.}
\end{figure}

Each map from the $H_0$ and $H_1$ sets were inpainted. An
example of a restored map is shown in Fig. \ref{fig:ex_simu_inp}.

We then apply the Hu\&Okamoto quadratic estimator on the full-sky, using
the fast spherical harmonic computations provided by the \hp package, with  
a multipole cut
$\lmax=2000$, since there is no statistical gain in going to higher
values given the noise level.
The "observed" temperature spectrum that enters the QE filter is
estimated for each map, which is a way of "absorbing" the residual spectrum deformation after
the mask restoration. 
We did not adopt the Bielewicz's trick and therefore insert the theoretical unlensed
spectrum into the numerator of the QE filter. Given the resolution and noise, we also analytically computed the
$\None$ and $\Ntwo$ terms using the fiducial lensing power spectrum.

\begin{figure}[htbp]
  \centering
  \includegraphics[width=\linewidth]{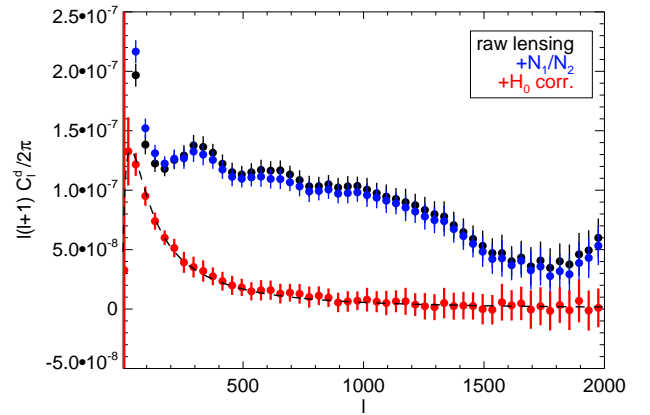}
  \caption{\label{fig:cldd_inp}  Mean deflection spectra
    reconstructed by applying the \inp method to the lensed maps. "Raw lensing" denotes the
    spectrum reconstructed directly from the maps. In blue, we show the
    effect of correcting for the (known) analytical terms \None and \Ntwo. In red, one
    subtracts the Monte Carlo correction obtained from the set of
    unlensed maps. The dashed line is the true input spectrum. All
    points are assigned an error bar corresponding to the sample variance of
    each map within our Monte Carlo set.
  }
\end{figure}

The bias size can then be estimated in either the $H_0$ or $H_1$ simulations. In the former case, one directly measures a
bias, after \Nz subtraction, that by construction, does not depend on the potential
field. In the latter case, one can estimate the bias with respect to the
fiducial model, after the $\Nz, \None$, and $\Ntwo$ corrections have been
applied, that can grab some extra contributions. We wish to check
the robustness of our correction of the lensing field, by estimating
\Nmc i the $H_0$ set.
The inpainting process is expected to induce a non-zero lensing
coupled bias, since it cannot accurately restore the lensed signal statistic up to the four-point
correlation function into the masked region. 
However, this effect can be \emph{effectively} accounted for by
introducing an \fsky factor to correct 
for the lack of power caused by the un-restored lensing modes within the
mask, so that the QE variance is given by
\begin{eqnarray}
\label{eq:inpvar}
  \ell (\ell + 1)\E{{\hat \phi_{\ell m}}^*{\hat \phi_{\ell' m'}}} &=& \delta_{\ell\ell'}\delta_{mm'}(\fsky \Cld + N_\ell^{dd}),\nonumber \\
\text{where } N_\ell^{dd} &=& \Nlz+ \fsky (\Nlone + \Nltwo) + \Nlmc, 
\end{eqnarray}
and the various contributions are shown in Fig.~\ref{fig:cldd_inp}.

The bias of the estimator is quite important but corrected for by
using the
\textit{unlensed} simulations. This means that the correction does not couple significantly to the
lensing spectrum. It however still introduces \textit{systematic} uncertainties
related to our limited knowledge of the instrument.
This motivates the development of an alternative method, which
completely avoids the masking issue, to the detriment of introducing some new technicalities.

\section{The local approach: \mpa}
\label{sec:mpa}
We start with a simple question. How do you derive the power
spectrum from a vector of sampled data? There are two approaches:
\begin{enumerate}
\item One solution is to perform a one step fast Fourier transform
  (FFT), which produces many modes that can then be averaged (binned) later on.
\item if the low frequency power spectrum is not required, a solution is to slice your sample into bunches, apodize each one, perform the individual FFTs, and take their mean. 
\end{enumerate}
Which approach is better? It was found that using the second one, with
overlapping segments of data (by $\simeq 50\%$) provides the nicest
(binned) estimates \citep{dsp}. This is known as the Welsh
periodogram.
That then happens when some portion of the data is missing? 
In the first case, one tries to \textit{correct} for the
gaps, possibly by restoring a mixture that has the correct statistical
properties given some prior of the signal, \ie by performing an inpainting. 
It is much simpler in the multi-bunch case, where one rejects chunks that overlaps 
with the gaps, an approach that is efficient as long as there are few of them
and they are largely contiguous. 

In the following, we apply these ideas to the case of data located on a
cut-sphere. We extend beyond the power spectrum estimation (which was largely studied
in \citet{das09}) and investigate whether this simple idea can be applied to CMB
lensing reconstruction, where the main "gap" is the Galactic
plane and the "bunches" are some tangent square planes.

We thus developed a pipeline that allows for a local reconstruction
of the CMB lensing in patches. This has the obvious advantage of avoiding the masked
regions and should therefore not introduce the large bias due to the
mask correlations that appears in a full-sky analysis. Furthermore, the noise
inhomogeneity, which adds a sizable contribution to the lensing
deflection, is also reduced by working locally. 

Working spatially also allows us to easily inspect the quality of the
data in different regions of the sky and therefore constrain the
experimental systematics. The natural flat-sky formalism that is
applied can be easily interpreted, and indicators, 
such as one for lensing isotropy, can be developed.

Statistically, after determining the Fourier complex coefficients for
each patch, we use the Hu-Okamoto quadratic estimator (QE)
described in \sect\ref{sec:ideal}. This has, by construction, a
minimum variance so there is no statistical loss in using this
approach. However, obviously, no scales below the patch Fourier size
 $\tfrac{2\pi}{L}$ can be reconstructed, hence we miss the low multipoles.

\subsection{Tiling the cut-sphere with patches}
The first unknown is the typical size ($L \times L$) of the
patches that one must use for lensing. It turns out to be a compromise between
several contradictory considerations: 
\begin{enumerate}
\item Lensing correlates modes over a few degree scale.
\item The Fourier modes that are to be
  reconstructed are located at harmonics of $k_0=\tfrac{2\pi}{L}$ in each
  $(k_x,k_y)$ Fourier direction. For $L=5\deg, 10\deg$ and $15 \deg$, respectively, this
  corresponds to $k_0=72,36,24$, which sets the grid spacing of the measured
  modes. To derive our final result in reasonably small
  multipole bins, we therefore chose to adopt a large $L$ value.
\item When projecting the data from the sphere onto the local tangent plane (using a gnomonic projection), we
wish to avoid too much distortion, which implies that we should not
use too large $L$ values. The classical $L \lesssim 20\deg$ flat-sky upper limit to the
flat sky approximation\citep{white} was derived from power spectra
considerations and is not necessarily valid for the four point
statistics we consider in lensing.
\item A last consideration is the efficiency of tiling a given cut-sky surface with square patches, which
causes them to be small. In addition, inspired by
the Welsh periodogram,  we seek a configuration where the patches
overlap by about 50\%, so that there is a clear interplay between the patch central positions and their sizes.
\end{enumerate}

These considerations suggests that patches of angular size $\simeq 10\deg$ with $\simeq
  50\%$ overlap are appropriate. Although it is a many-parameter system,
  we found that a simple solution is obtained with patches of angular
  size $L=10 \deg$ located at the centers of
a \hp $\ns=8$ map pixels. In this case, each pixel in the sphere
falls on average into $\simeq 1.8$ patches. We note that the tiling details do not impact
the final result, since we performed the same analysis on
$L=12\deg$ patches (which leads to a pixel on average falling into 2.6patches) and obtained very similar results.

\begin{figure}[htbp]
  \centering
  \includegraphics[width=.3\textwidth,angle=90]{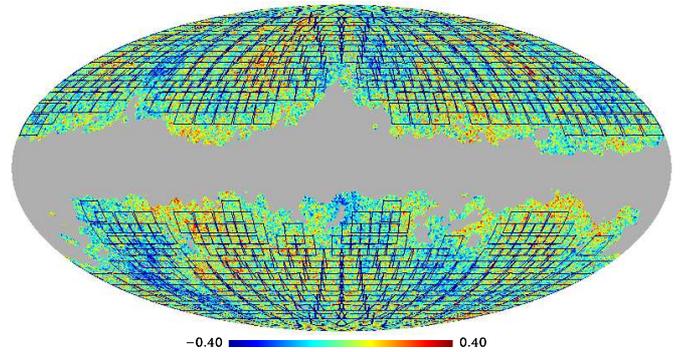}
  \caption{\label{fig:patch_pos} Example of the tiling of the map
    shown on Fig. \ref{fig:ex_simu}
    with overlapping $10\deg \times 10\deg$ patches, that do not
    intersect the Galactic masked region.}
\end{figure}

We then start with $12\ns^2=768$ patches.
Only patches that do not intersect the Galactic mask at all (the
reason being explained in the \texttt{prewhitening} section) are then kept,
which leaves $395$ of them, covering a fraction $f_\text{sky} = 55\%$
of the sky (as represented on Fig. \ref{fig:patch_pos}).
In this configuration, the overlap (the mean number of patches a point
of the sphere belongs to) is $\simeq 1.7$.

\subsection{Preparing the patches}

\paragraph{Local point-source inpainting.}
Before extracting the Fourier coefficients, we first need to
remove the bright
sources from the patches, which are a strong lensing contaminant.
This is performed by using the point-source mask and filling the
masked values by an inpainting algorithm.
We note that we use an (image) inpainting algorithm that differs from
the one described in \sect \ref{sec:inp}, because we wish each patch to
be treated independently of the others, which is not the case for \inp.
We chose a method that has
been designed and tuned for weak lensing  surveys
\texttt{FastLens}\footnote{\wwwfastlens}, which consists in minimizing
the sparsity of DCT (discrete cosine transform) coefficients for
$256\times256$ data blocks.

More precisely, we construct high resolution regular images from the
patches using bi-linear interpolation. 
The \texttt{FastLens} code is then run to fill in the point-source
masked areas. The inpainted values are then \textit{ back-projected onto the
sphere} to obtain again full-sky maps in which each sources belonging
to a patch have been filled.

Since the patches overlap, some filled sources sometimes belong to
several of them : we then use the inpainted
values from the patch whose center is the closest to the source, in order to avoid border effects.

\begin{figure}[htbp]
  \centering
  \subfigure[]{\includegraphics[width=.45\linewidth]{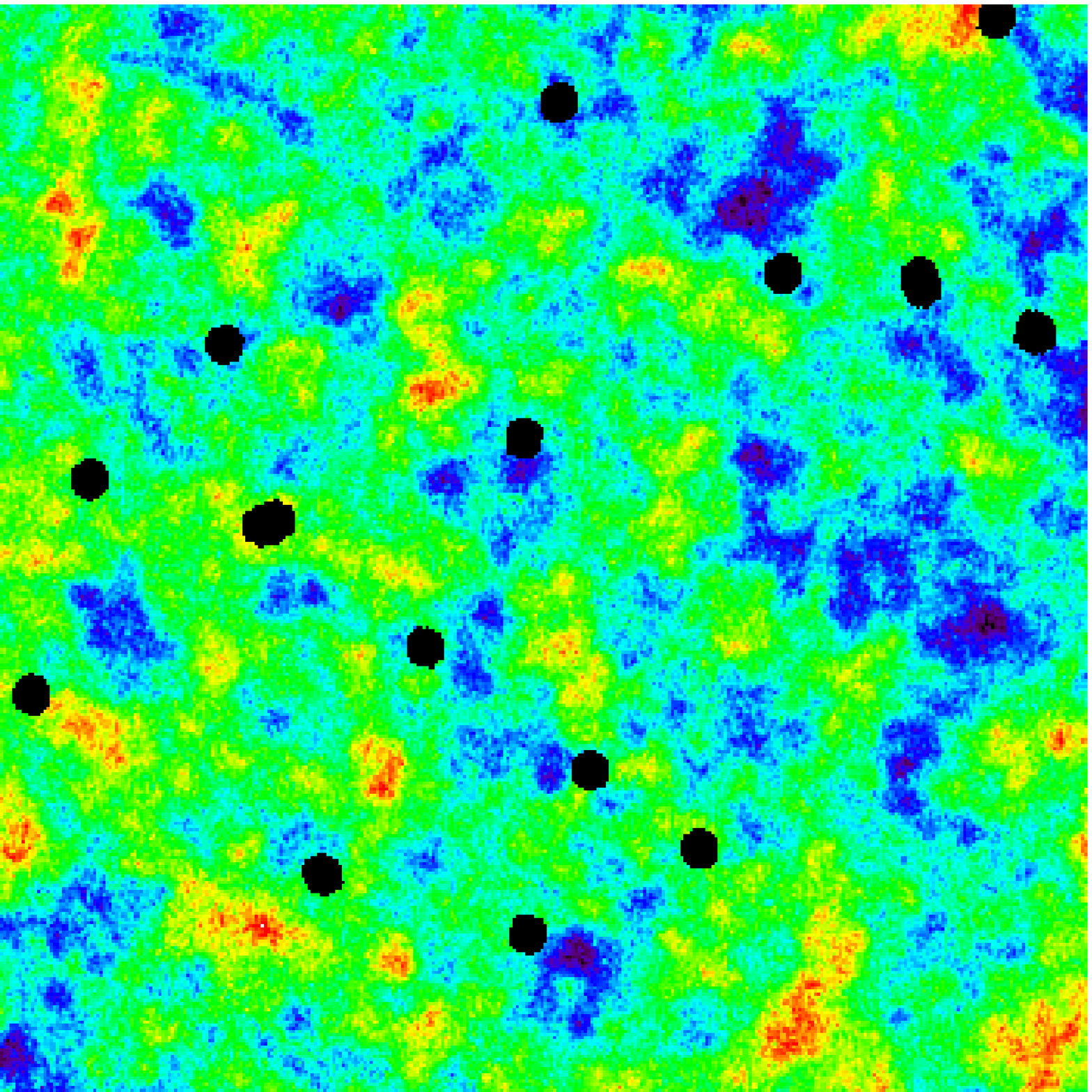}}
  \subfigure[]{\includegraphics[width=.45\linewidth]{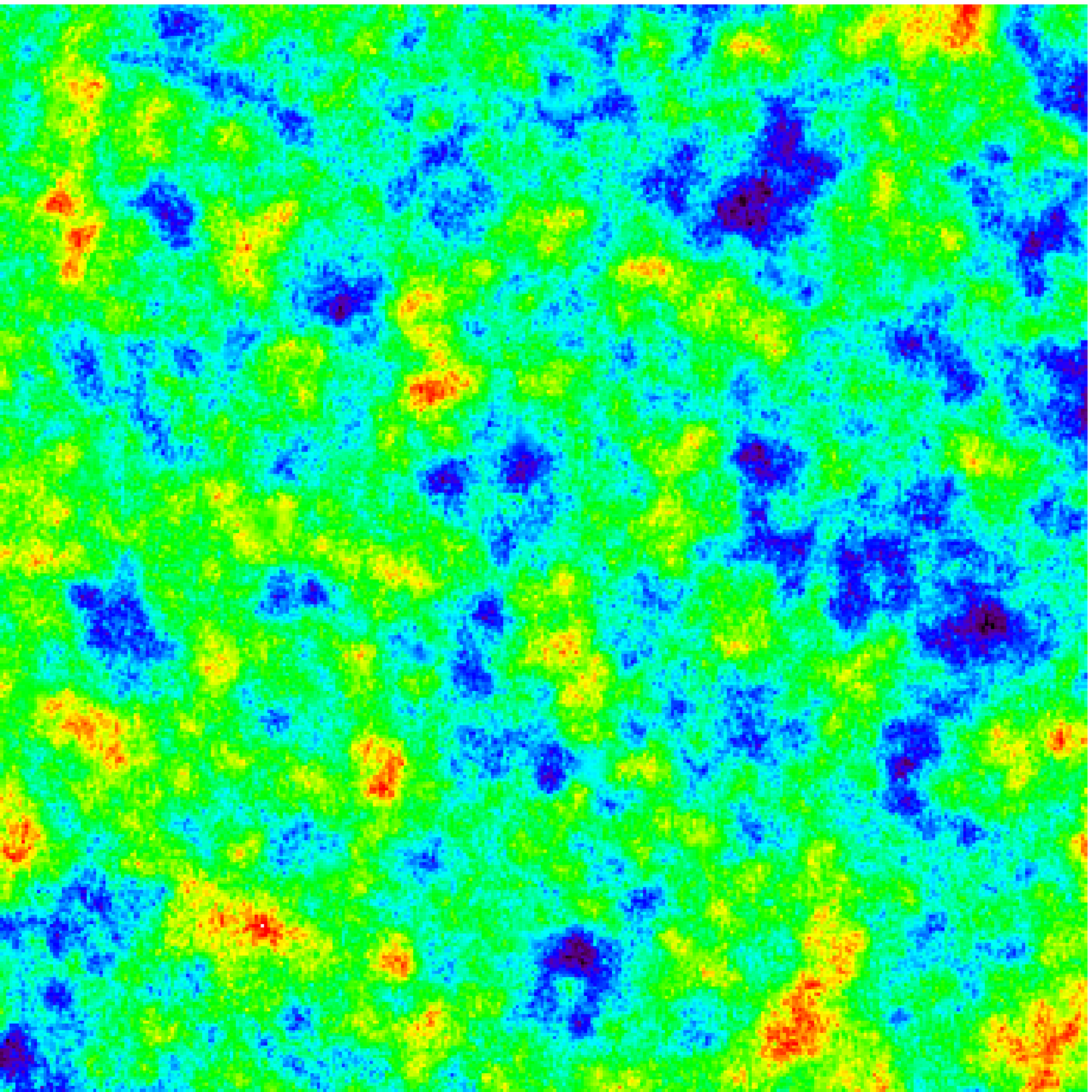}}
  \caption{\label{fig:patch_inp} 
    (a) Example of a $10\deg\times10\deg$ patch with masked sources
    in black. (b) Inpainting of the sources using the
    \texttt{FastLens} algorithm.
  }
\end{figure}

This procedure is applied to the full set of $H_0$ and $H_1$
simulations. An example is shown in Fig. \ref{fig:patch_inp}

\begin{figure}[htbp]
  \centering
  \subfigure[]{\includegraphics[width=\linewidth]{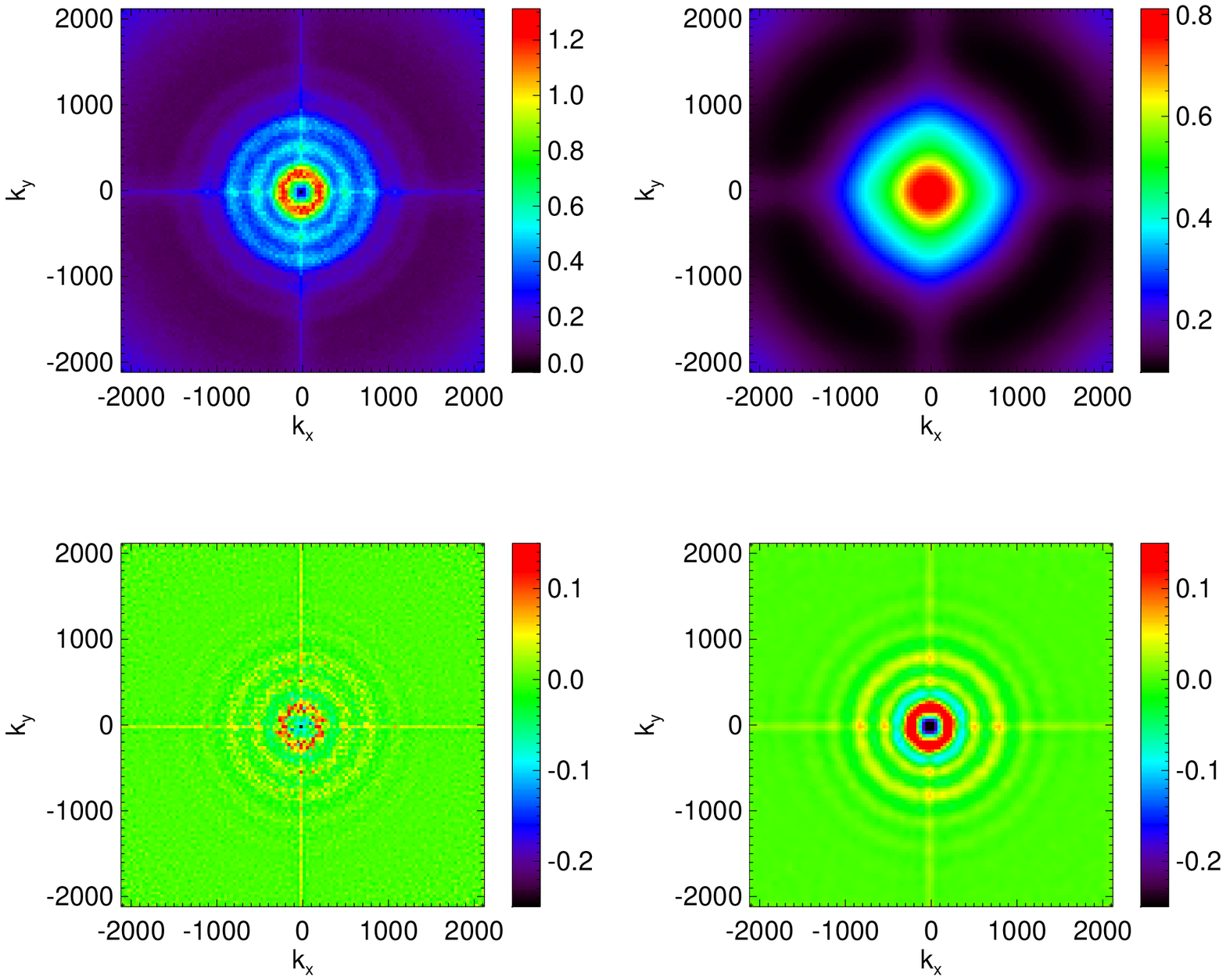}}
  \subfigure[]{\includegraphics[width=\linewidth]{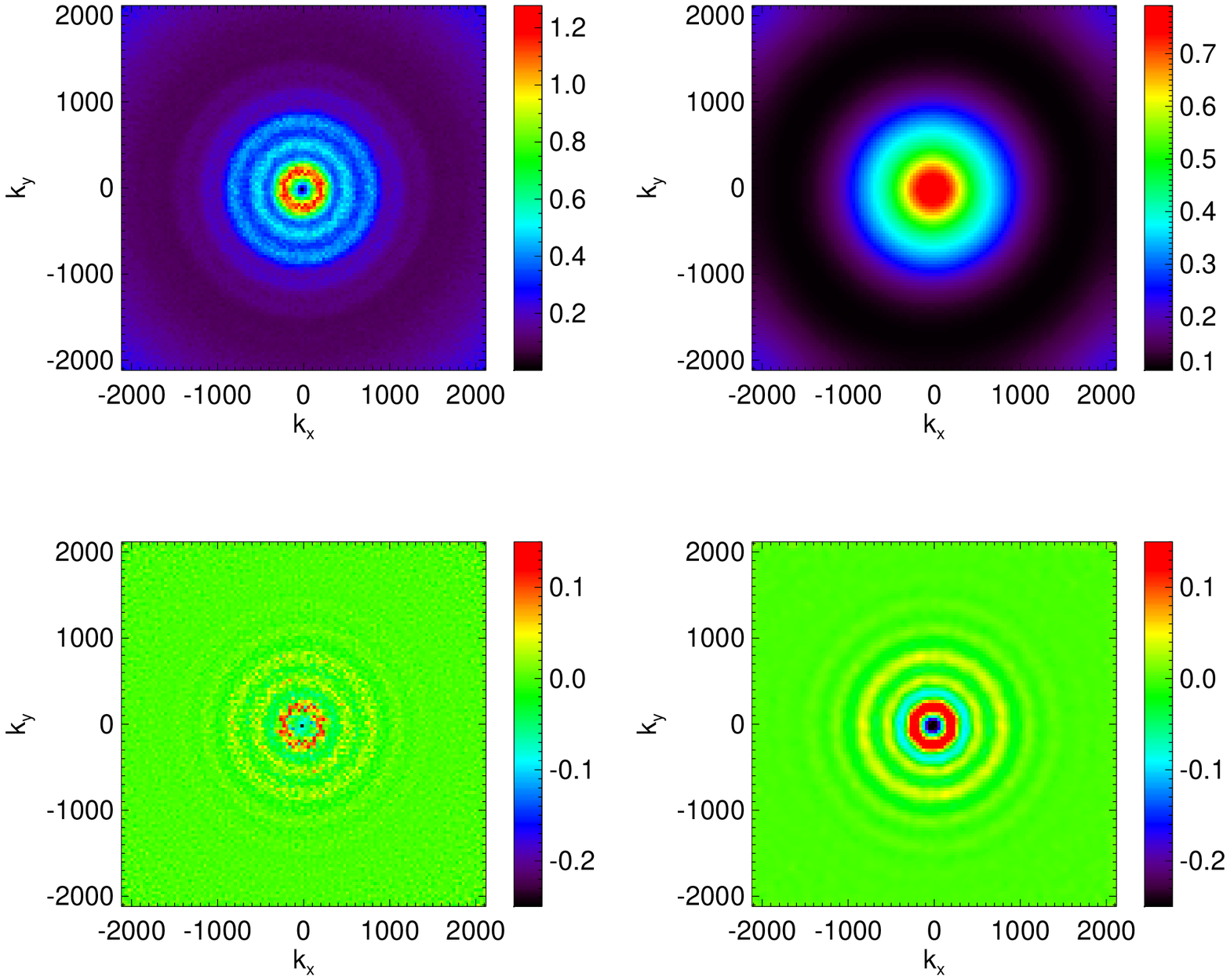}}
  \caption{\label{fig:ps_leakage} (a) Bi-dimensional Fourier spectra of
    one of our simulation at different scales. Upper-left: mean of
    the squared-amplitude of the Fourier coefficients for all patches
    \ie  $k^2 {\E{\norm{a_{{k_x},{k_y}}}^2}}$. An isotropic un-decimated
    wavelet transform ("a trous", see \eg \citet{starck}) is applied to the image and the results
    for scale one and two are shown in the bottom plots. The upper-right
    one corresponds to the smooth component. One
    notices a clear leakage along the null axes. (b) Same spectra
    but working on the prewhitened map and
    applying a Kaiser-Bessel $K_{0.5}$ window function. The leakage along the null
    axis has clearly disappeared.
  }
\end{figure}

\paragraph{Prewhitening and apodization.}
At this stage, we could again project the pixels onto patches and reconstruct
the Fourier coefficients (details in \sect \ref{subsec:fou}). However, we
note that the bi-dimensional temperature power-spectra obtained for these
patches exhibit a strong leakage along the null Fourier axis (Fig. \ref{fig:ps_leakage}(a) ).

We concluded that this leakage is due to that of the low $(k_x,k_y) $
modes which, for the CMB signal, have the stronger amplitudes, and  originates from the
side-lobes of the implicit $10\deg\times10\deg$ top-hat window used.
Instead of using some anisotropic filtering (the lensing itself being a
source of anisotropy), we can correct this by
\textit{prewhitening} the map and applying an explicit window.

\begin{figure}[htbp]
  \centering
  \subfigure[]{\includegraphics[width=.49\linewidth]{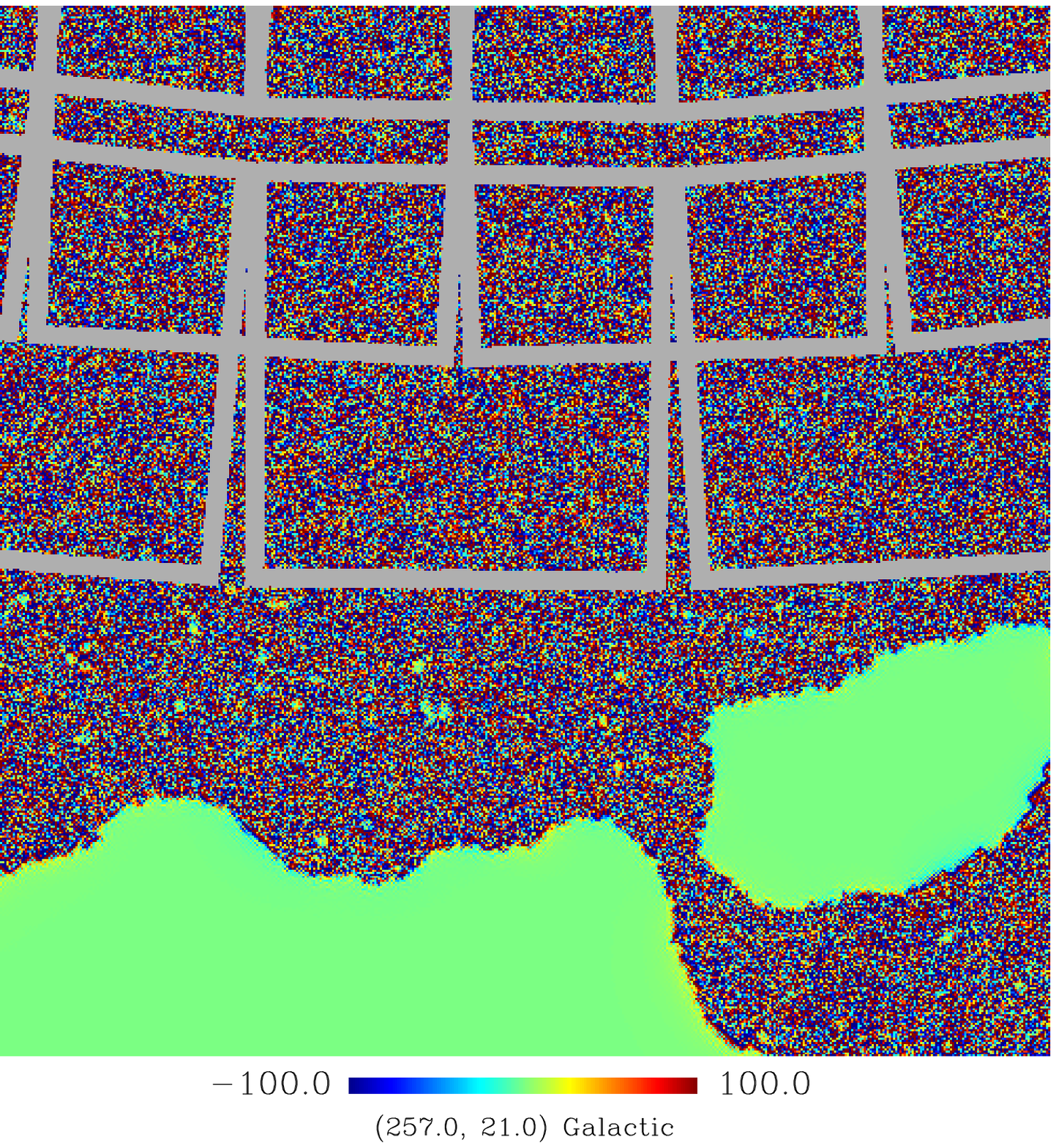}}
  \subfigure[]{\includegraphics[width=.49\linewidth]{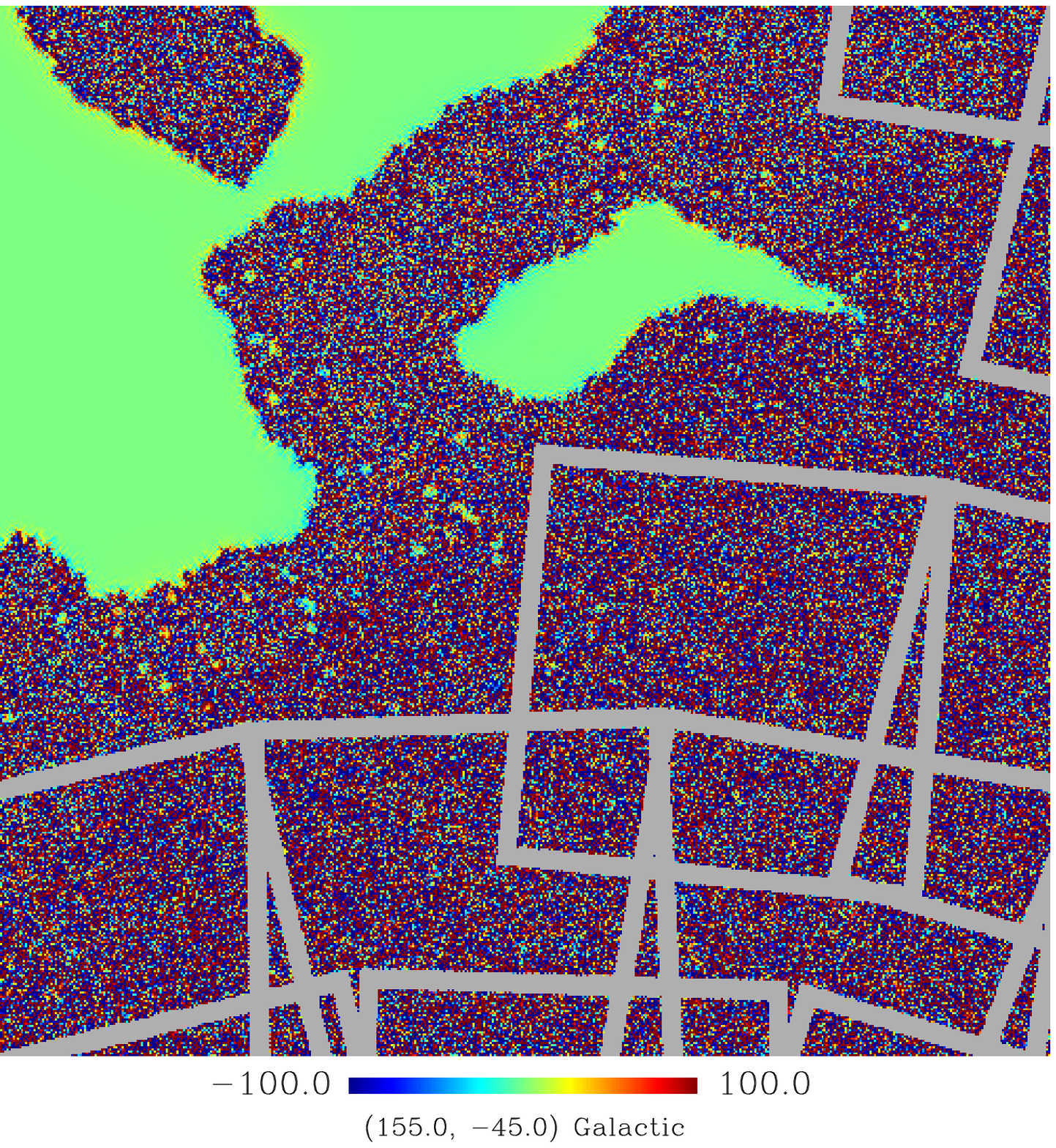}}
  \caption{\label{fig:prewhit} Examples of the $10\deg \times 10\deg$ \mpa tiling (in gray)
    of a \textit{prewhitened} map, around the borders of the Galactic mask (in
    green) in two regions of the sky (Galactic coordinates given in degrees). 
    One can discern some very local ringing around the boundaries of the mask
    and some point-sources located outside the patches and that had
    therefore not been previously inpainted.
}
\end{figure}

Prewhitening is a standard means of achieving comparably sized Fourier
coefficients \citep[\eg][]{das09}.
Since we are interested in a range up to
$\ell \simeq 2000$ (\ie which is not too far into the CMB damping tail), 
we need to approximately scale the
spectrum by $\ell^2$. We therefore simply multiply the
spherical harmonic coefficients of the map by $\ell$, and return to
direct space. In this process, the Galactic values are replaced by zero's, 
which results in some ringing
around the edges of the mask, which is fortunately damped by the instrument main
lobe. This is why we only used patches that do not intersect the
mask edges at all, since in practice, they are placed far enough away from the
mask frontier. Two illustrative examples are shown in Figure \ref{fig:prewhit}.

We note that the prewhitening
procedure was applied to both the "signal" maps ($H_1$ set) and
the MC-correction ones ($H_0$), so that if it still had a sizable impact
on the lensing reconstruction it would have biased the final lensing
estimate. Anticipating a result that is be presented later
(Fig. \ref{fig:cldd_patch}), the
$H_0$ correction is then found to be small, which validates
\textit{a-posteriori} the prewhitening procedure 
(and actually the entire procedure) is harmless to lensing.

From now on we work with these full-sky prewhitened maps for which the harmonic coefficients
are of similar order.

Rather than using for each patch the implicit top-hat window
(which has large side-lobes), we
apply an explicit window in the direct space. We work with
the family of Kaiser-Bessel functions \citep{kaiser}, which allow us to
vary simply the side-to-main lobe ratio, and that is still close to
the optimal solution of energy concentration provided by the
discrete prolate spheroidal sequence \citep{slepian,das09}.

Each value in the $L\times L$ size patch is therefore multiplied by
\begin{eqnarray}
  \label{eq:kaiser}
  W_\alpha(x,y)& =& W^{(1)}_\alpha(x) W^{(1)}_\alpha(y)\\ 
  W^{(1)}_\alpha(x)&=&\dfrac{1}{I_0(\pi\alpha)}I_0\left(\pi\alpha \sqrt{1-\left( x/\tfrac{L}{2}\right)^2}\right),
\end{eqnarray}
where $I_0$ denotes the zero-th order modified Bessel function of the first kind.

The Fourier transform of these windows is \footnote{In \refeq{kaiserK},
  a complex continuation is to be understood for low $k_x$ values}
\begin{eqnarray}
  \label{eq:kaiserK}
  \tilde W_\alpha(k_x,k_y)&=& \tilde W^{(1)}_\alpha(k_x) \tilde
  W^{(1)}_\alpha(k_y) \nonumber \\
  \tilde W^{(1)}_\alpha(k_x)&=& \dfrac{L}{I_0(\alpha \pi)} \sin_\text{c}\left(\sqrt{\left(\dfrac{k_x L}{2}\right)^2-(\alpha \pi)^2}\right),
\end{eqnarray}
which exhibits how the windows shrinks with $\alpha$ when comparing it
to the tophat window in Fourier space: $\tilde W^{(1)}(k_x)=\sin_\text{c}(\tfrac{k_x L}{2})$.

\begin{figure}[htbp]
  \centering
  \subfigure[]{\includegraphics[width=.8\linewidth]{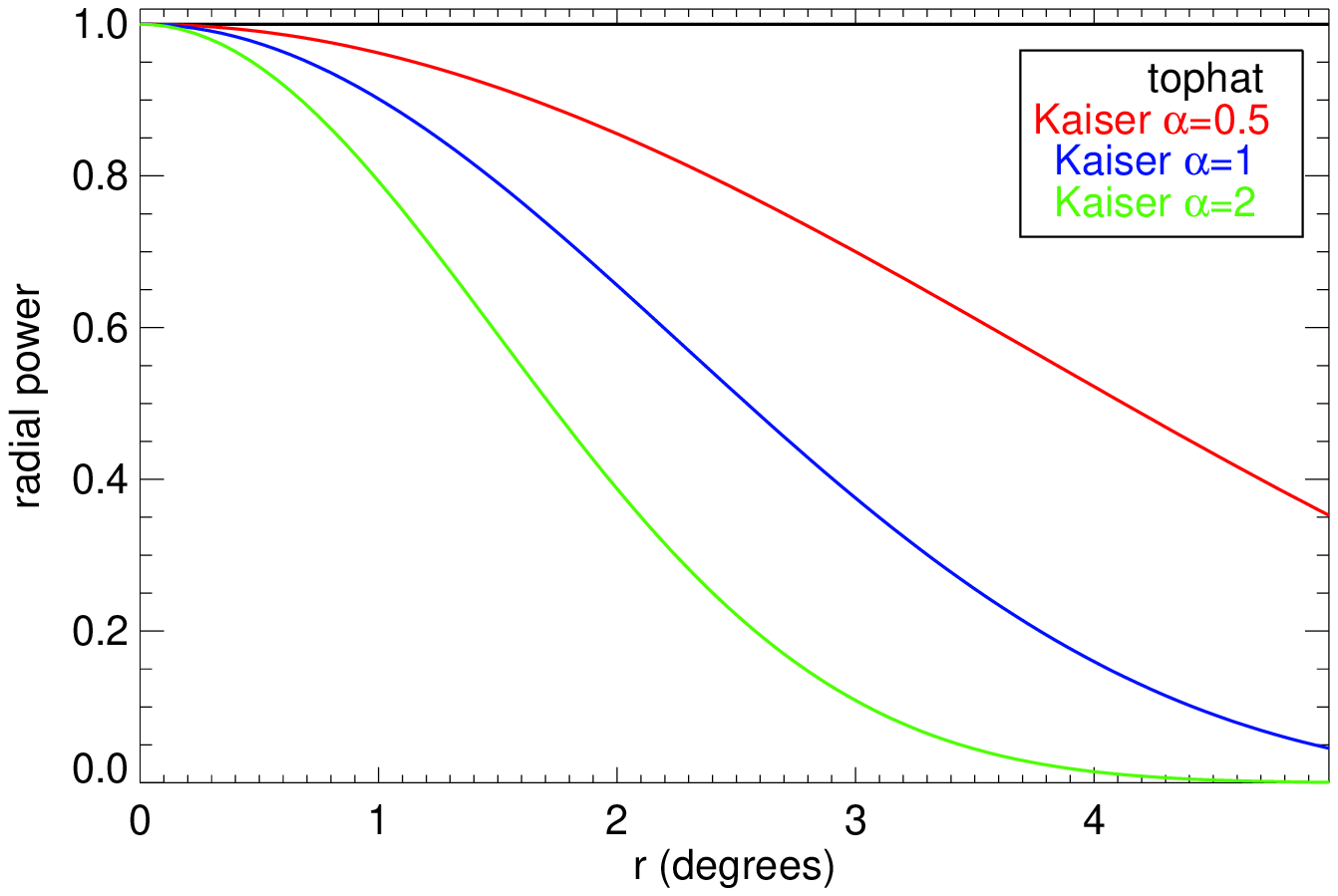}}
  \subfigure[]{\includegraphics[width=.8\linewidth]{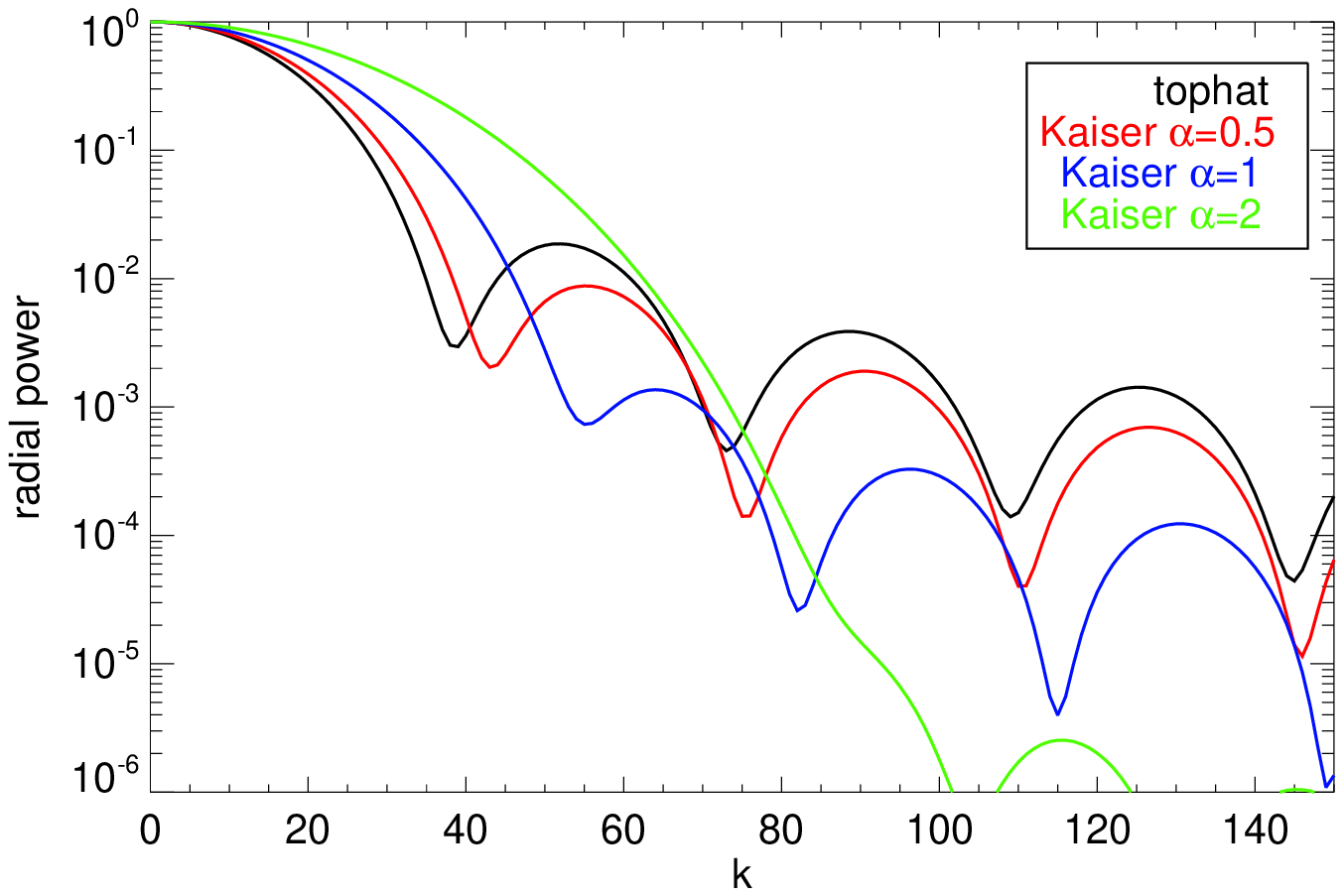}}
  \caption{\label{fig:kaiser} 
    Radial power of the Kaiser-Bessel 2D windows with
    $\alpha=0.5,1,2$ for $L=10\deg$ in real (a) and Fourier
    (b) space. The top-hat result is also shown.
  }
\end{figure}

We compute numerically the radial power of these windows as :
\begin{eqnarray}
  \label{eq:radial}
  P_W (r)&=&\dfrac{1}{2\pi}\int_0^{2\pi} |W_\alpha(r \cos\phi,r \sin \phi)|^2 d\phi \nonumber \\
  P_{\tilde W} (k)&=& \dfrac{1}{2\pi}\int_0^{2\pi} |\tilde W_\alpha(k \cos\theta,k \sin \theta)|^2 d\theta
\end{eqnarray}
and show them for $\alpha=0.5,1,2$ on Fig.~\ref{fig:kaiser} in direct
and Fourier space.

As is well-known, diminishing the side-lobes always occurs at the price
of increasing the main lobe width (energy conservation). 
In the following, we describe how we attempted to keep $\alpha$ as small as
possible to keep the window strongly peaked since the QE considers the
products of modes are convolved by this window in Fourier space
(\sect \ref{sec:deflection_patch}).

We checked that after prewhitening and windowing with the
Kaiser($\alpha=0.5$) window,  
the power spectrum leakage disappears as is clear
on Fig.~\ref{fig:ps_leakage}(b). In the following, we therefore
use $W_{0.5}$ as an explicit apodization function.

The size of the window in Fourier space , Fig. \ref{fig:kaiser}(b),
fixes the binning.
For $W_{0.5}$ on a $L=10\deg$ square patch, we use a step of
$\Delta k=40$, starting from the first available Fourier mode $k_0=35$,

\subsection{Fourier series estimation of non-equispaced data}
\label{subsec:fou}
We project the prewhitened map onto the local patches, using a
gnomonic projection. The \hp pixel centers then fall onto an irregular
bi-dimensional grid. Can we still perform a spectral analysis ? 

For a $\ns=2048$ \hp map, the mean inter-pixel
separation is about $1.7\arcmin$, which is \textit{not} negligible
compared to the expected mean deflection of the CMB lensing ($\simeq
2.7 \arcmin$), so that an interpolation would induce some large effects.
To avoid this interpolation, we therefore developed
the so-called  "ACT" (Adaptive weight, Conjugate gradient, Toeplitz
matrices) algorithm, 
which allows us to \textit{fit} the complex Fourier coefficients from a
set of irregularly sampled data in a reasonable
time (see also \cite{nfft}). This method has been proposed for real
(1D) data and we generalized it to bi-dimensional data.
We give hereafter the main idea and discuss the technicalities in the Appendix.

We search for the least squares estimates of the $a_{k,h}$ (complex) Fourier coefficients
of our band-limited temperature signal, in the series expansion
\begin{equation}
  T(x,y) = \sum_{k=-M_x}^{M_x}\sum_{h=-M_y}^{M_y} a_{k,h}e^{2i\pi k \frac{x}{L}}e^{2i\pi h \frac{y}{L}}.
\end{equation}
In the general case, the brute force inversion of the normal
equations is prohibitive, but
the method takes advantage of the peculiar structure of the Fourier series 
decomposition to perform operations very efficiently.
In our case, we manage to determine the $120\times120$ coefficients
of the $\simeq 120~000$ data values contained in a $10\deg
\times10\deg$ patch of an \hp $\ns=2048$
map, in about one minute on a single core computer.

The tool we developed, named \ft,  has been compared to a standard FFT method, when fitting
(actually solving) $N \times N$ points with $N \times N$ unknowns on a \textit{regular} grid:
our results agree to within machine precision.

This tool opens the road to local analyses of projected spherical
maps, which are plagued by interpolation issues. It has been
used successfully in computing the spectra and full CMB bi-spectra in \citet{pires}.

\subsection{Local power spectra estimates}

After running the \ft tool, we have an estimate of the complex
Fourier coefficients per patch, at wave-vector $\vec k$, located on the
regular grid \refeq{grid}.

By computing the squared-amplitude map, one can study the 2D local
power spectrum on the sky, and even though the cosmic variance is large,
detect potential experimental problems. By taking the mean of the
power spectra for all the patches, one can check for the CMB field isotropy in a simple way.

\begin{figure}[htbp]
  \centering
  \includegraphics[width=\linewidth]{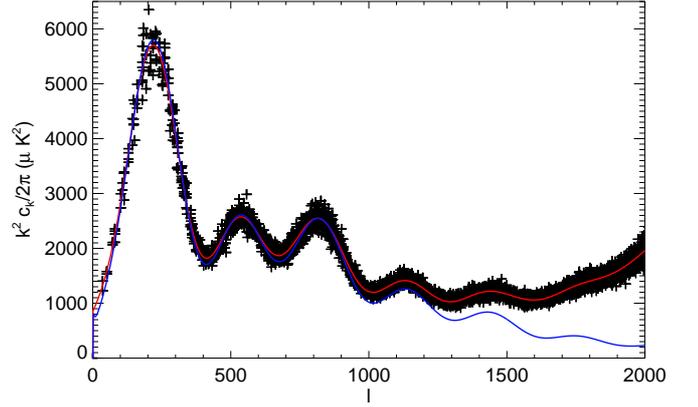}
  \caption{\label{fig:psd} Example of a 1D power
    spectrum used in the lensing estimator. The points are obtained from
    the bi-dimensional spectrum (as in Fig. \ref{fig:ps_leakage}(b)),
    deconvolved from the beam, and represented \wrt to \kmod. They are
    fitted to the smooth function in red. Also shown in blue is the
    fiducial model.
  }
\end{figure}

By plotting the values, \wrt \kmod (\ie assuming isotropy), one
constructs a 1D power spectrum which is equivalent to the famous \cl but
for the non-integer values \kmod given by \refeq{grid}.
One has a powerful local power-spectrum estimator that solves the issues of
masking that is well-suited to jackknife tests. To get a full determination of the spectrum, one
would still have to study the window function, as in
\citet{master} and \citet{das09}, but we do not actually need it for the
lensing reconstruction since it relies on the \textit{observed} spectrum.
We first need to deconvolve the maps from the main lobe, which is a trivial
operation in the Fourier space for a Gaussian shape, and obtain some smooth
spectrum. This is obtained by taking the mean power spectrum of the
de-convolved Fourier patches, and fitting the coefficients of a
generic smooth function to all the data points as explained in 
\citet{fitcls}. The result is shown for one of our simulation, in
Fig.~\ref{fig:psd}.

\subsection{Local deflection estimates}
 \label{sec:deflection_patch}
After determining the complex Fourier coefficients for each patch, and the
observed/true \cl's, we may apply the Hu\&Okamoto flat-sky
estimator to obtain the (noisy) potential maps in the Fourier domain,
\refeq{esti}. But before going on to the lensing spectrum estimate, we need to review
the noise of the estimator on an \textit{apodized} patch since the standard QE has not been derived in this case. 

The quadratic term $\E{T(\vec k_1)T(\vec K-\vec k_1)}$ upon
which the estimator in \refeq{esti} is build, is affected in a non-trivial way by the
apodization procedure. \citet[][Appendix B]{metcalf} show that it introduces:
\begin{enumerate}
\item An "aliasing" effect due to the overlap of the windows in Fourier
  space that affects only the low $\ell$'s modes.
\item Some complicated "smoothing" of the lensing potential. 
\end{enumerate}

We do not try to build an optimal estimator from the elaborate
expression.
We note instead that the apodization process
scales the lensing Gaussian noise merely by a constant factor, that can be
understood in the following way.

On the basis of \citet{master} and \citet{efsta}, one can show that,
assuming that the window is well peaked in Fourier space such that the
spectrum does not vary too much over it, the two-point correlation
function of an apodized Gaussian field $T^{\textrm{apo}}$ along with
its variance can be approximated by
\begin{eqnarray}
\label{eq:apocalcula}
{\E{T^{\textrm{apo}}(\vec k_1)T^{\textrm{apo}}(\vec k_2)}} & \simeq &
(2\pi)^2 \delta(\vec k_1 + \vec k_2) w_2 C_{k_1} \\
\label{eq:apocalculb}
\textrm{var}(T^{\textrm{apo}}(\vec k_1)T^{\textrm{apo}}(\vec k_2)) &\simeq & 2 w_4 C_{k_1} C_{k_2} ,
\end{eqnarray}
where
\begin{equation}
  \label{eq:wi}
  w_i=\dfrac{1}{L^2} \int_{-L/2}^{L/2} \int_{L/2}^{L/2} dx dy W^i(x,y) 
\end{equation}
and $W(x,y)$ is the window function in direct space.

These approximations are accurate for large $k$ values
\citep{efsta}, which corresponds, given our window size to $k \gtrsim 100$.

In the following, we use windows normalized by $\sqrt{w_2}$, so that the
reconstructed power spectrum, the only entity that varies with
apodization in the filter/normalization of the QE, \refeq{esti2}, is mainly unchanged (\refeq{apocalcula}). We
checked for instance that applying a $W_{0.5}$ window would change the
lensing normalization $A_k$ by less than 1\%.

\begin{figure}[htbp]
  \centering
  \includegraphics[width=\linewidth]{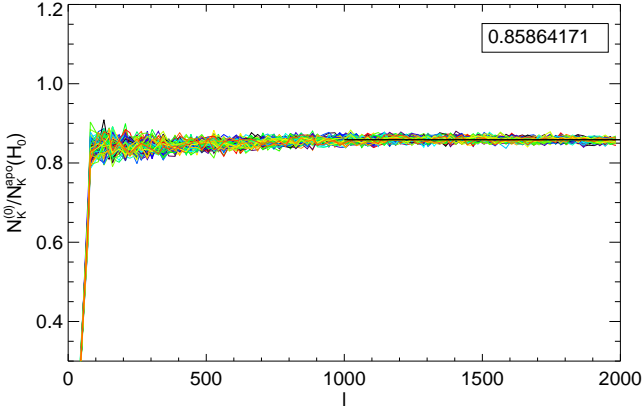}
  \caption{\label{fig:factor} Color plots showing the ratio of the
    non-apodized \Nz term \refeq{n0} to  
    the reconstructed lensing
    variance in the patches apodized by $W_{0.5}$ for each of the 100 maps
    of our unlensed ($H_0$) set. A constant term, whose value is depicted in the upper
    box, is fitted to the $\ell \ge 1000$ part.
  }
\end{figure}

The Gaussian noise in the apodized case can be written as the variance
in the QE applied to the apodized \emph{unlensed} sky:
\begin{equation}
 N_K^{(0, \textrm{apo})} = K^2 {\E{\norm{\hat \phi^{\textrm{unlens, apo}}(\vec K)}^2}}.
\end{equation}
Substituting \refeq{apocalculb} into this equation and using the normalized window, one obtains  
\begin{equation}
\label{nl0final}
N_K^{(0, \textrm{apo})} \simeq \frac{w_4}{(w_2)^2} N_K^{(0)}.
\end{equation}
What is the accuracy of this approximation?
We ran the QE on the unlensed simulations, computed the mean lensing spectrum
and compared it to the ideal case given by \refeq{n0}. The result in Fig. \ref{fig:factor} shows
that the ratio is indeed reasonably flat for values of $\ell \gtrsim
100$. Fitting the mean value of this ratio in the high $\ell$ region
gives a result very close to the analytical value $\tfrac{(w_2)^2}{w_4}= 0.859$.

We performed the exercise for several windows, including the often used
Hanning one \citep[\eg][]{dsp}, and give our results in Table
\ref{tab:apod}) . They all show excellent agreement with the simple
$\tfrac{(w_2)^2}{w_4}$ scaling factor.

\begin{table}[htbp]
  \centering
  \begin{tabular}{|c|c|c|}
    \hline
    window & $\tfrac{(w_2)^2}{w_4}$ & MC-$H_0$ \\
    \hline
    Top-hat & 1 & 1 \\
    Kaiser($\alpha=0.5$) & 0.859 & 0.859 \\
    Kaiser($\alpha=1$) & 0.502 & 0.505 \\
    Kaiser($\alpha=2$) & 0.248 & 0.249 \\
    Hanning & 0.264 & 0.265 \\
    \hline
  \end{tabular}
  \caption{ Comparison of the lensing excess Gaussian noise due to various
    apodization windows. $\tfrac{(w_2)^2}{w_4}$ is the analytical
    approximation, whereas the last column gives the measured ratio
    from the unlensed Monte Carlo's, obtained as on
    Fig. \ref{fig:factor}.
  \label{tab:apod}}
\end{table}

This leads us to rewrite the QE total covariance \refeq{varphi} in the apodized case, as
\begin{equation}
  \begin{split}
  K^2 {\E{{\hat \phi(\vec K)}^*{\hat \phi(\vec K')}}} &= (2\pi)^2
  \delta(\vec K - \vec K') \\
  & \cdot \dfrac{w_4}{(w_2)^2} [C_K^d +\Nkz+\Nkone+\Nktwo + \Nkmc]
  \end{split}
\end{equation}
and propose the simple estimator for an apodized patch
\begin{equation}
\label{eq:estiapo}
K^2\widehat{C_K^\phi} = \frac{(w_2)^2}{w_4} \intphi K^2 |\hat
\phi^{\textrm{apo}}(\vec K)|^2 - (N_K^{(0)}+ N_K^{(1)}+\cdots)
\end{equation}
where the integral is performed in small size rings over
the discrete Fourier grid.

This \textit{proposed} estimator, \refeq{estiapo}, is then tested
on our $H_1$ simulations in order to assess its bias/variance.
We note that the value of the scale factor does not need to be
known very precisely. What matters is that the \textit{same} factor is
used in the data (here $H_1$) and the Monte Carlo correction (here $H_0$).

We now have all in hand to compute the deflection power-spectra. This
is performed for each map of our Monte Carlo set, in the following way:

\begin{enumerate}
\item From the Fourier coefficients obtained on each patch, we form the
  2D-Fourier potential map $\hat \phi$ using \refeq{esti}. The
  normalization (and \Nz term) 
  is computed in the standard way, using
  Eqs. (\ref{eq:n0}) and (\ref{eq:esti2}) and the true/observed spectrum
  as given in Fig.~\ref{fig:psd}.
\item For each patch, we form the noise-corrected deflection power map
  of $f K^2 \norm{\hat \phi_K}^2 -N_K^{(0)}$
  where $f=0.86$ in our case.
\item We accumulate the 395 power maps and evaluate their mean and variance.
\item We compute the inverse-variance weighted average in rings of
  constant $\Delta K=40$ width (starting at $K_0=35$).
  The binned values are reported at the mean of the different
  modes $K_{i,j}$ locations within the ring.
\end{enumerate}

We compute these spectra in the $H_1$ set, which still have noise
contributions from the $\None$, $\Ntwo$ and $\Nmc$ terms.
The $\Nmc$ term (the "bias") is taken from the $H_0$ simulations as the
mean difference from 0 of the reconstructed spectra following the same procedure.

\begin{figure}[htbp]
  \centering
  \includegraphics[width=\linewidth]{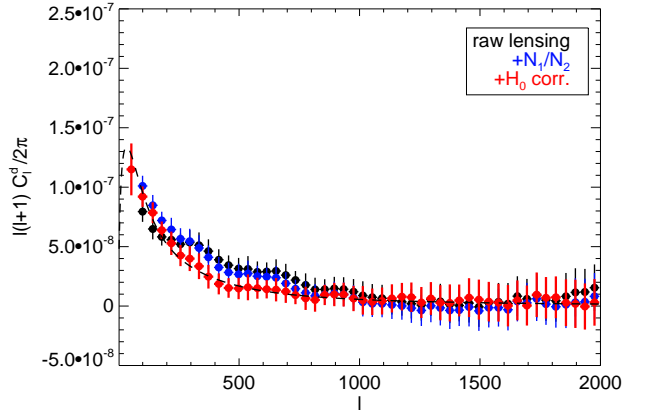}
  \caption{\label{fig:cldd_patch}  Mean deflection spectra
    reconstructed by the \mpa method from the lensed maps. "Raw lensing" denotes the
    spectrum measured directly on the maps as described in the text
    without performing any bias correction. In blue, we show the
    effect of subtracting the (known) analytical terms $\None$ and $\Ntwo$. In red, one
    accounts for the Monte Carlo correction obtained for the
    set of all 
    unlensed maps. The dashed line is the fiducial input spectrum. All
    points are assigned an error bar corresponding to the variance in 
    the Monte Carlo simulations per sky map. The first bin for the raw-lensing
    estimate is located outside the plot (at a value of $3.5 \times
    10^{-7}$). The same range as on Fig. \ref{fig:cldd_inp} has been
    used for proper comparison.
  }
\end{figure}

The mean spectrum is shown in Fig. \ref{fig:cldd_patch}, where one sees the
various contributions.

The initial (raw) lensing spectrum estimate already has a very little bias, thanks to
our lack of use of any Galactic data and to a reduced local noise inhomogeneity.

We show that the \mpa method leads, after a small correction, to an un-biased
estimate of the deflection over the whole $\ell \in [75,2000]$ range.
The very first bin $[35,75]$ is also unbiased but receives a stronger correction
from the $H_0$ MC, which is due to the breakdown of the flat-sky limit
($\ell$ cannot be identified to \kmod in this case) and to 
the apodization window having an overlap integral
that extends to approximately 100 (see Fig.~\ref{fig:kaiser}(b)).

We note that unlensed simulations accurately correcting  
the lensed ones means that \textit{the full reconstruction process does
  not induce any (significant) couplings to the underlying lensing potential}.
The variance in the estimator is discussed in the next part.

\begin{figure}[htbp]
  \centering
  \includegraphics[width=.9\linewidth]{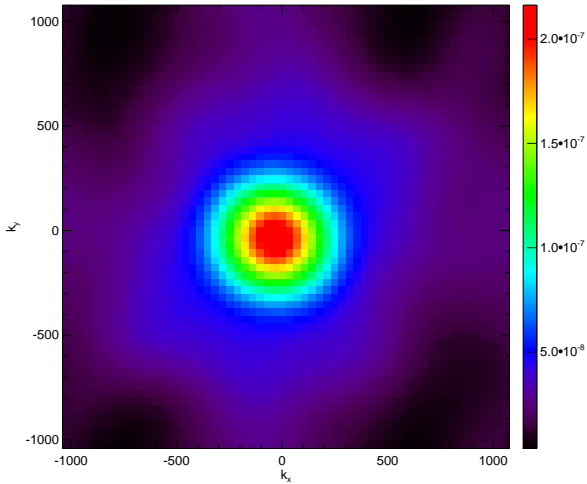}
  \caption{\label{fig:ck2d_lensing} Example of an isotropy check of the
    lensing potential. For one of our lensed maps, we evaluate the "raw
    lensing" estimator (\ie that does not include any MC correction)
    for each patch, and take their mean. We represent the power map
    $k^4 \hat C_k^\phi$ after $\Nz$ subtraction, and smoothed by an "a trous" transform, as in the
    upper right part of Fig. \ref{fig:ps_leakage}(b).
  }
\end{figure}

Finally, we note that working on patches has a number of other benefits: 

\begin{enumerate}
\item We derive bi-dimensional maps of the lensing potential, so that
  as in Fig.~\ref{fig:ps_leakage}, we can more easily test the deflection field
  isotropy. An example is show in Fig.~\ref{fig:ck2d_lensing}.
\item For each patch, we can check for unexpected systematic error
  effects that would provide an excessive lensing signal (as missed sources).
\item With knowledge of all sources of noise, one can apply a Wiener
  filter to each patch to reconstruct the
lensing potential maps that can then be cross-correlated to other
cosmological probes of the matter, such as Galactic weak lensing
(cosmic Shear) or cosmic infrared background, 
which are only generally measured over a small region of the sky.
\end{enumerate}

\section{Comparison of the methods}
\label{sec:results}

\begin{figure}[htbp]
  \centering
  \subfigure[]{\includegraphics[width=\linewidth]{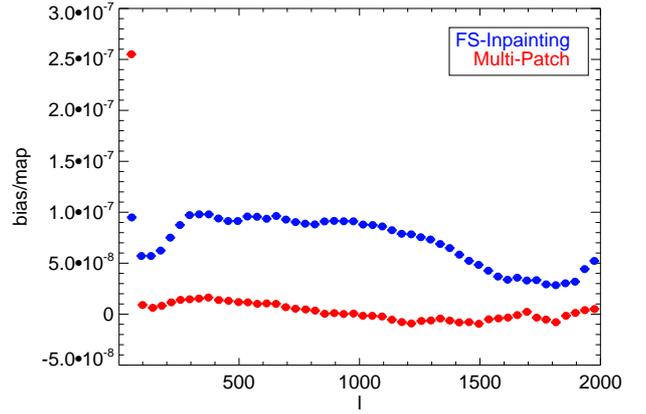}}
  \subfigure[]{\includegraphics[width=\linewidth]{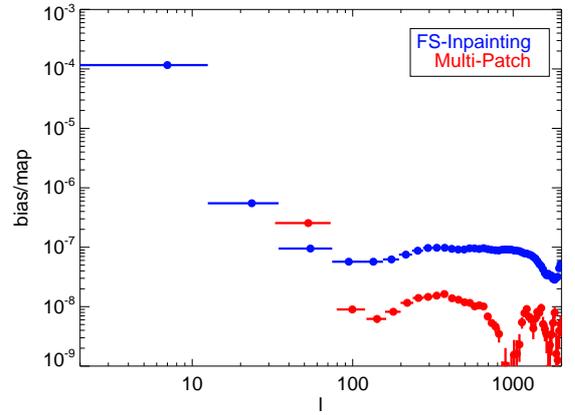}}
  \caption{\label{fig:biais} (a) Bias of the methods computed from our
    simulations as the mean of the spurious deflection power on the
    unlensed set. For \inp, it corresponds to $\Nlmc/\fsky$ in
    \refeq{inpvar}, 
    whereas for \mpa, it is the result of applying the modified QE to
    $H_0$ as described in \sect \ref{sec:deflection_patch}. (b)
    Same plot with a log scale to
    emphasize the low $\ell$'s. No mode below 35 is available to the \mpa method.
  }
\end{figure}

In Fig.~\ref{fig:biais}, we compare the bias, in each $\ell$ bin, 
of the \inp and \mpa methods. They were obtained using \textit{unlensed} simulations, and
were indeed shown to correct the deflection power measure in lensed maps.
We recall that the binning that we used 
starts at $\ell=35$ (first \mpa accessible mode for a $10\deg\times 10 \deg$
patches) and has a width $\Delta \ell=40$.  Two extra bins were 
added to \inp $[2,13]$ and $[14,34]$.

\begin{figure}[htbp]
  \centering
  \subfigure[]{\includegraphics[width=\linewidth]{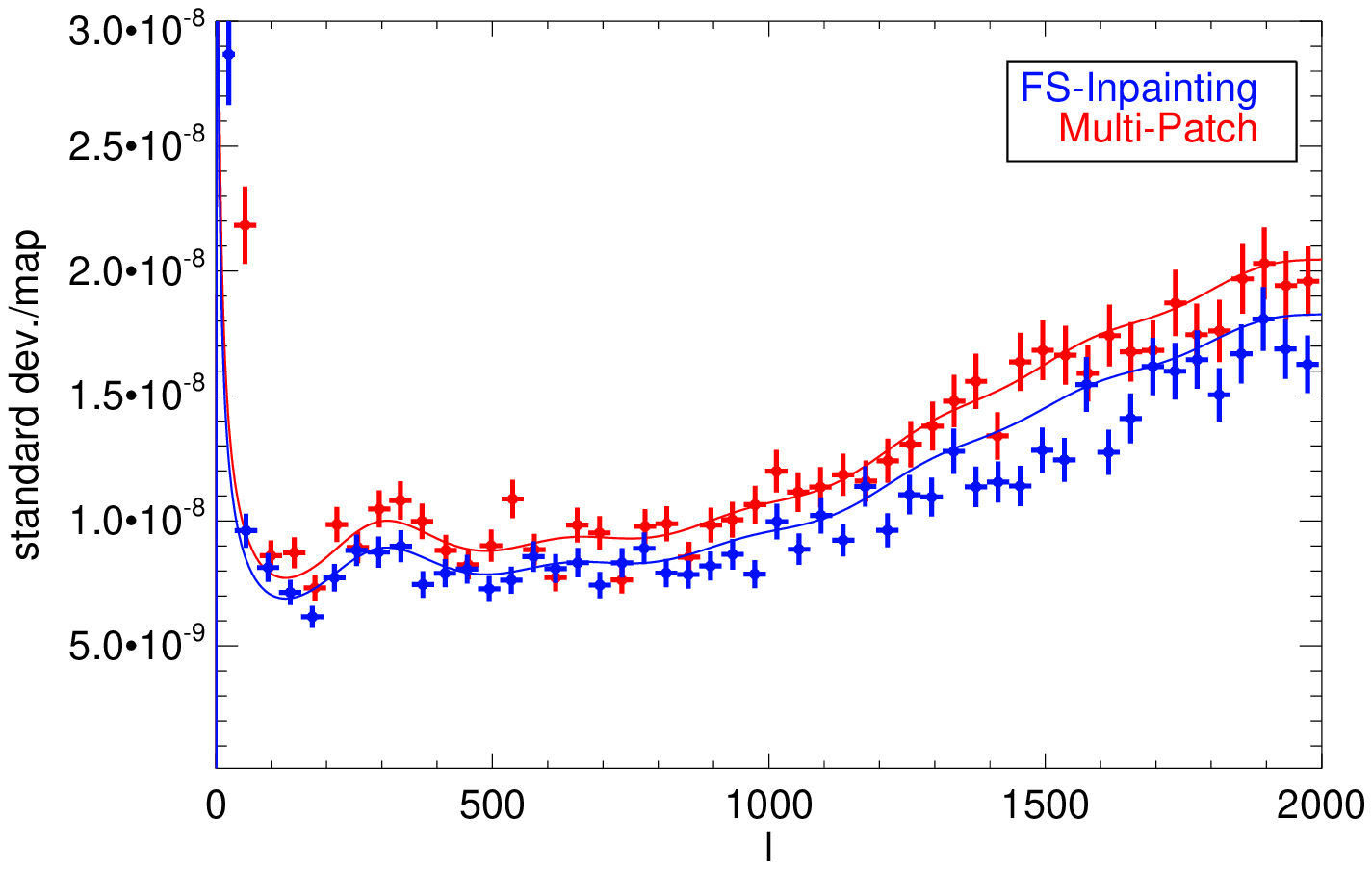}}
  \subfigure[]{\includegraphics[width=\linewidth]{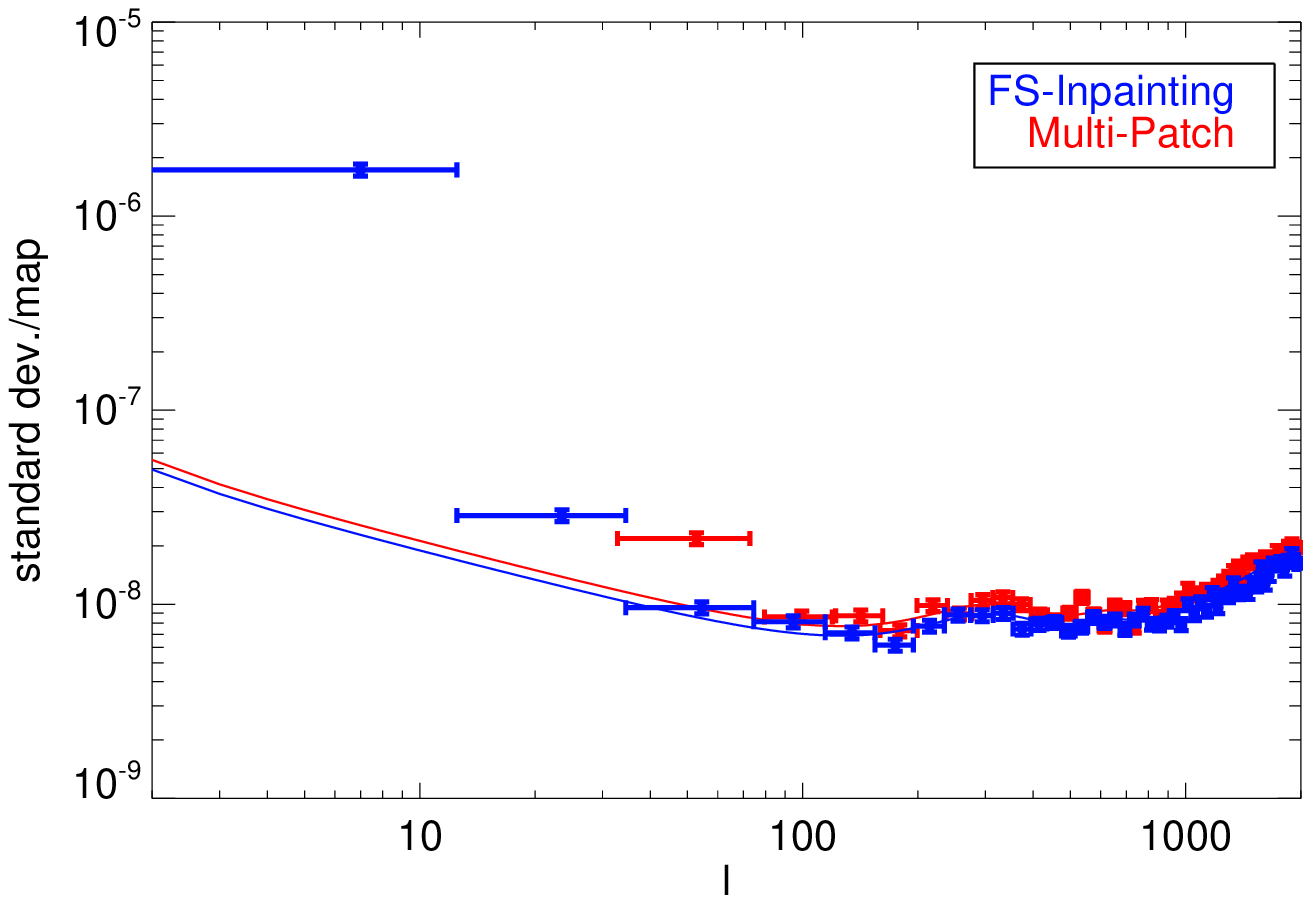}}
  \caption{\label{fig:var} (a) Standard deviation among the methods computed from our
    simulations, as the spread in each bin of the lensing estimators
    for the 100 $H_1$ set. (b) Same plot with a log scale to emphasize low $\ell$'s.
  }
\end{figure}

The standard deviation in each bin from the $H_1$ set is shown in
Fig. \ref{fig:var}. The Fisher error estimate for the QE 
\begin{equation}
\label{eq:fisher}
\sigma_\ell = \dfrac{\Cld + \Nlz}{\sqrt{\ell\Delta \ell \fsky}}
\end{equation}
is also depicted. We recall that the \mpa method has a lower
sky coverage ($\fsky=0.55$ ) than the inpainted one ($\fsky=0.69$ )
owing to the procedure for tiling the cut-sky.

From these plots, it appears that
\begin{itemize}
\item The \mpa approach clearly shows less bias, except for the very
  first bin $[35,75]$. It cannot reconstruct modes below these values.
\item For large $\ell \gtrsim 100$, both estimators follow the naive
  Fisher error estimates, the \inp one having a slightly smaller 
  variance than for \mpa, owing to the larger sky coverage.
\end{itemize}

We feel that the small relative statistical loss ($\simeq 10\%$) of the \mpa
method is largely compensated for by the gain on systematic errors. 
As explained in \sect \ref{sec:mpa}, the reconstruction can be
controlled and visually inspected at each step, but most importantly \textit{the
systematic errors due to the bias of a necessarily imperfect simulation are minimized}. 

We therefore advocate the use of an hybrid method consisting in the \mpa approach for
$\ell\gtrsim 100$ and \inp for lower $\ell$'s.

\section*{Conclusion}

We have investigated two methods to reconstruct the lensing-deflection
power spectrum from \pl-like CMB frequency maps, using a large Galactic
cut and including some strong noise inhomogeneity.
The first one, \inp, which was previously presented in \citet{papier1}, is still
found to be efficient in this 
extreme configuration as long as one corrects for a large yet
lensing-independent bias using Monte Carlo simulations. 
We have developed a well-suited method to deal with large masks, based
on tiling the cut-sky with 
$10\deg\times 10 \deg$ patches and performing
local analyses. This has required us to solve some problems related to non-periodic
boundary conditions and Fourier coefficients determination for
irregularly sampled 2D data.
For this purpose, we have developed the \ft tool, which allows the fast 
\textit{exact} fitting of the Fourier series coefficients in irregularly
sampled 2D data. This is a valuable tool for other analyses that
require a high level of precision at the spatial location.

Both methods have been demonstrated for realistic
\pl-like simulations of the 217 \ghz CMB channel. It was found that the
\mpa approach has a very low bias in the whole $100 \le \ell \le 2000$
range, thanks to the avoidance of the Galactic plane, and lower local noise
inhomogeneity. It allows us at each step to check for experimental systematic
errors and perform local images of both the temperature and deflection
bi-dimensional power spectra.
Its final variance is only marginally larger than a 
full-sky method, and could be improved by a smarter strategy for tiling
the cut-sky sphere. 
The final result is insensitive to the precise
position of the patches and of their overlap.
To perform some cosmological fits using the reconstructed spectrum,
the inter-bin correlation would still have to be measured accurately,
and included in the likelihood, since we have measured some
$(15 \pm10)\%$ correlation level, with our $\Delta \ell=40$
binning. This requires a large number of simulations ($\simeq 1000$).

In the $\ell \le 100$ range, we advocated using the \inp method, which 
provides the minimal variance estimate in the cut sky.
Since our simulations did not include a
  Galactic contaminant, this boundary could slightly shift.
  However, using our $30\%$ Galactic mask, we checked by adding a
  simulated Galactic component to our maps, that its net effect on the
  reconstructed deflection power was extremely negligible (on real
  data, some template would be subtracted).

These results open the road to measuring CMB lensing directly in \pl-like
CMB maps, without even performing a component separation of the foregrounds. 
This is not exactly true for the \inp method, where one must have 
clean boundaries at the mask frontier.
However, this can be obtained by a
simple template subtraction measured in the high frequency channels. For the \mpa
method, one can still perform the analysis without "un-dusting" the
map, by choosing appropriate CMB-dominated patches; 
we checked in simulations that a sub-dominant amount of dust
contamination does not affect the lensing deflection spectrum.

It is not our goal to come to a decision on whether a component
separation method is a more accurate means of lensing reconstruction, since it is
an area that remains under active development. We note however that
the statistical gain offered by using a larger fraction of the sky can
be counterbalanced by a higher \Nz term due to a larger (combined)
lobe of the instrument or a higher final noise level (see  \refeq{fisher}).
Adding the high/low frequency channels will also "bring back" some 
additional infrared/radio sources that need
to be masked out, lowering the final statistical gain of the combined map.

Working directly on intensity maps allows us to perform various sanity
tests by checking the consistency between the reconstructions from
different frequency maps, which is a way of assessing the robustness of
the estimate against either experimental uncertainties or physical
contamination -- as from possible SZ-lensing or unresolved
radio-source-lensing correlations.
The reconstructions of each frequency maps can also be
minimum-variance combined, which offers a robust reconstruction that allows
for a high level of systematic control. 
Finally, this is a well-suited approach to cross-correlation studies with
other mass tracers, by selecting frequencies that are unaffected by
contaminants that may induce extra correlations. For instance, one
may wish to use only
frequencies over 100 \ghz (with negligible unresolved radio-source
contamination), while studying correlations with external radio
surveys, 
or below 217\ghz for a CIB-lensing correlation estimate. 

\begin{acknowledgements}
We acknowledge the use of \texttt{CAMB} \citep{camb}, \hp \citep{hp},
\texttt{LensPix} \citep{lenspix}
, \texttt{MRS} \citep{mrs}, \texttt{FastLens} \citep{fastlens}, and
\texttt{FuturCMB2} \citep{futurcmb2} packages.  
We thank Simon Prunet for the precise derivation of the effective
number of degrees of freedom in \cite{master} and
Martin Reinecke for some dedicated \hp \texttt{C++} developments.
\end{acknowledgements}

\newpage
\appendix
\section*{Appendix: \ft tool}

In order to not lose accuracy in determining the Fourier
coefficients from a sample of irregularly sampled points, 
we developed a tool for \textit{fitting} these
coefficients in a reasonable time.

We start with the 1D case, where we have implemented
the "second generation" algorithm proposed in \citet{feichtinger}. 

We define $f$ to be a function sampled on any support $\{t_i\}$.
In a given interval $(0,T)$, the function can be expanded into a
Fourier series
\[ f(t) = \sum_{k=-\infty}^{\infty} a_ke^{2i\pi k \frac{t}{T}}.\]
Assuming that it has a band-limited spectrum, so that we can limit the
number of Fourier modes to
\[ f(t) = \sum_{k=-M}^{M} a_ke^{2i\pi k \frac{t}{T}},\]
the problem is to determine the $a_k$ coefficients given the sampled
values $f_i$.

We consider the reduced $u=\tfrac{t}{T}$ variable. If the number
of samples $u_i$ is N, one writes the N equations
\[ f_i = \sum_{k=-M}^{M} a_ke^{2i\pi k u_i}.\]
This is a linear system of N equations with 2M+1 unknowns. The well-
known normal equations obtained from least squares minimization are 
\[(G^TG)X= G^TF,\]
where $F$ is the column vector of sampled values and $G$ is the matrix
with elements $g_{kl}=e^{2i\pi k u_l}$.
The solution is in general computationally heavy using standard methods.

Here, the interesting point is that the generic term of the system is
of the type:
\[ T_{kl}= \sum_{j=1}^{N}e^{-2i\pi(k-l)u_j},\]
which is a Toeplitz matrix.
One solves the system using the conjugate-gradient algorithm (the matrix is
Hermitian and positive), which consists in performing successive matrix-vector
products. One then pads the Toeplitz matrix with zeros to obtain a
circulant matrix, since the product of a circulant matrix with a
vector can be computed efficiently using an FFT. 

We extended this method to the 2D case using the 
formalism of the Kronecker products of matrices and the properties of
the separability of FFTs.
This allows for the fast determination of the $a_{kh}$ coefficients in the
Fourier expansion
\[ f(x,y) = \sum_{k=-M_x}^{M_x}\sum_{h=-M_y}^{M_y} a_{kh}e^{2i\pi k
  \frac{x}{T_x}}e^{2i\pi k \frac{ty}{T_y}} .\]
In our case, a $10\times10\deg$ patch from an \hp $\ns=2048$ map,
contains about $120 000$ irregularly sampled values. 

The modes are harmonics of $\Delta k_x=\Delta
k_y=\tfrac{2\pi}{L}\simeq 35$, so we need to determine $120\times 120$
(half is negative) of them to obtain all modes below $l_{max}=2000$.
This is performed in about one minute on a single core. The conjugate
gradient converges in about seven iterations, without using any special
pre-conditioner, so we did not add the adaptive weight scheme.

\bibliographystyle{aa}
\bibliography{refs}

\end{document}